\documentclass[12pt,preprint]{aastex}
\usepackage{graphicx}
\usepackage{subfigure}

\newcommand{\myemail}{t.evans@physics.usyd.edu.au}

\slugcomment{Accepted for publication in The Astrophysical Journal}

\shorttitle{ASTROPHYSICAL FALSE POSITIVES IN GROUND-BASED SURVEYS}
\shortauthors{EVANS \& SACKETT}

\begin{document}

\title{An \emph{a priori} investigation of astrophysical false positives in ground-based transiting planet surveys}

\author{Tom M. Evans\altaffilmark{}}
\affil{Research School of Astronomy and Astrophysics, The Australian National University, Mount Stromlo Observatory, Cotter Road, Weston Creek, ACT 2611, Australia}
\email{\myemail}

\author{Penny D. Sackett\altaffilmark{}}
\affil{Research School of Astronomy and Astrophysics, The Australian National University, Mount Stromlo Observatory, Cotter Road, Weston Creek, ACT 2611, Australia}
\email{}

\begin{abstract}
Astrophysical false positives due to stellar eclipsing binaries pose one of the greatest challenges to ground-based surveys for transiting Hot Jupiters. We have used known properties of multiple star systems and Hot Jupiter systems to predict, \emph{a priori}, the number of such false detections and the number of genuine planet detections recovered in two hypothetical but realistic ground-based transit surveys targeting fields close to the galactic plane ($b\sim 10^\circ$): a shallow survey covering a magnitude range $10<V<13$,  and a deep survey covering a magnitude range $15<V<19$. Our results are consistent with the commonly-reported experience of false detections outnumbering planet detections by a factor of $\sim$10 in shallow surveys, while in our synthetic deep survey we find $\sim$1-2 false detections for every planet detection. We characterize the eclipsing binary configurations that are most likely to cause false detections and find that they can be divided into three main types: (i)~two dwarfs undergoing grazing transits, (ii)~two dwarfs undergoing low-latitude transits in which one component has a substantially smaller radius than the other, and (iii)~two eclipsing dwarfs blended with one or more physically unassociated foreground stars. We also predict that a significant fraction of Hot Jupiter detections are blended with the light from other stars, showing that care must be taken to identify the presence of any unresolved neighbors in order to obtain accurate estimates of planetary radii.  This issue is likely to extend to terrestrial planet candidates in the CoRoT and Kepler transit surveys, for which neighbors of much fainter relative brightness will be important.

\end{abstract}


\section{Introduction}   \label{sec1}

The majority of extrasolar planets that have been discovered by ground-based transit surveys to date are so-called ``Hot Jupiters'', with radii and masses similar to Jupiter and orbital periods less than $10$ days~(d). Ground-based transit surveys can be divided into two main types: (i) shallow surveys that use telescopes with apertures $\sim$10cm to monitor $\sim$50\,deg${}^2$ field of views (FOVs) over a magnitude range $V \lesssim 13$ \citep[eg.][]{2004ApJ...613L.153A,2007ApJ...656..552B,2007MNRAS.379..773S}, and (ii) deep surveys that use $\sim$1m telescopes to monitor $\sim$1\,deg${}^2$ FOVs over a magnitude range $V \lesssim 19$ \citep[eg.][]{2003AcA....53..133U,2009IAUS..253..333B,2009AJ....137.4368B}.  

Most transit searches have adopted the shallow survey approach, largely because the required equipment is less expensive and more readily available than that needed to conduct a deep survey. Another appeal of shallow surveys is that $\sim$1m class telescopes can be used to perform follow-up spectroscopy on their bright targets, with signal-to-noise~ratios~(S/N) that are sufficiently high to detect radial velocity variations caused by an unseen Jupiter-mass companion. The fainter targets of deep surveys, on the other hand, require follow-up spectroscopy to be performed on $\sim$6-10m telescopes in order to achieve the same level of precision. Certainly, spectroscopic investigations of planetary atmospheres are only currently possible for the bright systems monitored by the shallow surveys \citep[eg.][]{2002ApJ...568..377C,2007Natur.447..183K}. 

Nevertheless, shallow surveys do have a number of disadvantages compared to deep surveys. Perhaps most significantly, late-type main sequence stars are much rarer over the magnitude range covered by shallow surveys than they are in deeper surveys. More stars must therefore be monitored overall in a shallow survey than in a deep survey to have the same probability of observing a Hot Jupiter transit a late-type main sequence parent star. 

Another drawback of ground-based shallow surveys is the lower spatial resolutions that they achieve, typically limited by point-spread functions (PSFs) with full-width-at-half-maximum~(FWHM) sizes~$\sim$20$''$, whereas deep surveys typically have much smaller PSFs of~$\sim$2$''$. Spatial resolution is important in transit surveys, because blending with physically unassociated stars can dilute any actual planetary signal or cause a transit by an eclipsing binary~(EcB) system to be mistaken for a planetary system. Although the number of stars per square degree of sky is much higher at the faint magnitudes covered by deep surveys, we find that the difference in PSF sizes is the dominant effect, making blending more prevalent in shallow surveys~(Table~\ref{tab:nblends} and \S{\ref{sec:embf}}).

The planetary yield of any photometric transit survey is far outweighed by the number of detected EcBs \citep[eg.][]{2008AJ....135..649W,2008AJ....135..850D}. Generally, the light curve produced by an EcB is easy to distinguish from one produced by a Hot Jupiter transiting its parent star. The transit durations are typically longer and the primary eclipse depths are typically deeper for EcBs than they are for transiting Hot Jupiters. Also, because both components of an EcB are luminous, a detectable secondary eclipse is usually produced. Unless both components of the EcB have very similar radii and luminosities, the secondary eclipse will not have the same depth as the primary eclipse, immediately revealing the non-planetary nature of the signal. 

While most EcBs can be identified using one or more such diagnostics, in certain cases an EcB can mimic the photometric signal produced by a transiting Hot Jupiter. There are three main situations in which this can occur: (i) a late-type dwarf star eclipsing a larger, brighter dwarf star; (ii) two dwarfs undergoing grazing eclipses; (iii) two dwarfs eclipsing each other blended with the light of one or more unresolved stars. Meanwhile, detached EcBs with a giant component do not exist for periods less than $10$ days due to Roche lobe overflow \citep{2004A&A...421..241S}, and hence are not able to mimic the signal of a transiting Hot Jupiter. 

It has not been uncommon for the rate of false planetary detections due to EcBs to be over an order of magnitude higher than the rate of true planet detections \citep[eg.][]{2006ApJ...644.1237O}. The identification and rejection of such objects can be far from trivial and poses a significant challenge to searches for transiting planets. 

It is instructive to consider the yields of transit surveys in terms of what is already known about the properties of multiple star systems and Hot Jupiter systems from published empirical studies. \cite{2008ApJ...686.1302B} have developed a methodology to predict, \emph{a priori}, the number of planetary detections that a user-specified survey will make. Such estimates are helpful both in designing surveys and in evaluating the effectiveness of existing surveys. On the other hand, \cite{2003ApJ...593L.125B} made \emph{a priori} estimates of the rate of false detections in two synthetic transit surveys, one based on the shallow STARE survey and the other based on the then-planned Kepler space-based survey. A bottom-up understanding of the types of EcBs that are likely to cause false detections and the rates at which they occur is valuable both for interpreting the objects detected by transit surveys and for refining follow-up approaches used to reject non-planetary signals.

In this paper, we describe a method analogous to those of \cite{2008ApJ...686.1302B} and \cite{2003ApJ...593L.125B} to estimate the expected rates of false detections and Hot Jupiter detections in two synthetic surveys: (i) a ground-based shallow survey; and (ii) a ground-based deep survey. For each synthetic survey, the types of EcB configurations that are predicted to cause false detections are examined in some detail. These results are then compared with the reported experience of real transit surveys surveys, and the general implications for ground-based transit surveys are discussed. 

Primarily, our study builds upon that of Brown (2003), by making more rigorous a number of the simplifying assumptions contained in that earlier work. For instance, unlike Brown, we explore the potential for genuine Hot Jupiter detections to be combined with faint blends, which can affect the estimated planetary radius (\S\ref{sec:configs}), whereas he only investigated unblended Hot Jupiter detections. We also do not ignore EcB secondary eclipses and use a more realistic S/N criterion in our definition of what constitutes a detection (\S\ref{sec:crit}).  We have also attempted to account for the possibility that blended neighbors fainter than the ``formal'' upper magnitude limit of the survey can contribute in a non-negligible way to the observed photometric signals of stars within the formal magnitude range (\S\ref{sec:embf}). Furthermore, Brown only considered single-star blends in his calculations, but we have allowed for multiple-star blends (\S\ref{sec:phifb}). Our experience with Lupus-TR-3b in the SuperLupus survey \citep{2009IAUS..253..333B,2009AJ....137.4368B}, which we describe in \S\ref{sec2}, suggests that such cases might not be especially rare.

\section{Astrophysical false positives} \label{sec2}

Survey teams often report that false detections due to EcBs outnumber true planet detections by over an order of magnitude \citep[eg.][]{2006ApJ...644.1237O}. It should be noted that reports from shallow surveys dominate the literature, and this ratio may in general be lower for deeper surveys, as has been the experience of the SuperLupus survey \citep{2009IAUS..253..333B,2009AJ....137.4368B}. In any case, astrophysical false positives are a ubiquitous feature of any transit survey and the process of identifying and rejecting them has proven to be a time-consuming and, in many cases, subtle task.

Hot Jupiter transit signals can be imitated by: (i)~a late-type main sequence star eclipsing a larger main sequence star, (ii)~two main sequence stars undergoing grazing eclipses, or (iii)~two eclipsing main sequence stars blended with one or more other stars inside a photometric confusion radius, determined by either the instrumental PSF or the CCD pixel scale. For this reason, survey teams must follow up all promising candidates with additional observations to rule out the possibility that the photometric signal is caused by an EcB. Such observations include:
\begin{itemize}
  \item Multi-color photometry: Apart from high-order stellar limb-darkening effects, planetary transit signals are color independent. Signals due to EcBs, on the other hand, typically have a stronger color dependence that can be detected by making photometric measurements in more than one passband. 
  \item High spatial resolution imaging: Often a candidate that was originally presumed to be a single star will be revealed to consist of two or more fainter stars by higher resolution imaging. By monitoring the flux changes of each component separately, it is sometimes possible to identify astrophysical false positives caused by blended EcBs by recognizing that one of the resolved components undergoes an eclipse that is too deep to be caused by a planetary transit.
  \item Spectral typing: Spectroscopy can be used to identify the spectral type and luminosity class of the star that is being transited. If the star turns out to be a giant, then the photometric signal cannot be caused by a transiting planet since it would produce a flux change too small to be detected. Instead, the signal must be due to a blended EcB.
  \item Radial velocity measurements: For a solar-type star, variations in radial velocity caused by an orbiting stellar companion will be $\sim$10\,km\,s${}^{-1}$, whereas the oscillations will only be $\sim$0.1\,km$\,$s$
{}^{-1}$ for a Jupiter-mass companion. Radial velocity measurements with a precision of $\sim$1\,km$\,$s${}^{-1}$ can therefore be used to identify cases in which the companion is stellar. 
\end{itemize}

\cite{2003ASPC..294..409L}, \cite{2004AIPC..713..151C}, and \cite{2006ApJ...644.1237O,2007ApJ...662..658O} describe astrophysical false positives encountered as part of the Vulcan and TrES surveys, detailing how the non-planetary natures of candidate objects were revealed using one or more of the follow-up techniques listed above. 

An example of a more challenging false positive, and the resource-intensive processes that were employed to reject it, has been detailed by \cite{2005ApJ...621.1061M}. In this case, no color dependence was observed in the eclipse and initial spectral typing showed that the primary, GSC 01944-02289, was an F dwarf. Radial velocity measurements indicated that the companion had a mass of $\sim$30$\,M_J$. All of these observations were consistent with a brown dwarf transiting the primary. However, careful measurement of asymmetries in the spectral lines and extensive modelling of possible blend scenarios ultimately showed that the observed signal was caused by a faint EcB blended with another unresolved star rather than a transiting brown dwarf.

The experience of the SuperLupus survey \citep{2009IAUS..253..333B,2009AJ....137.4368B} has also reinforced how difficult it can be to determine the precise nature of candidates, especially for the faint objects targeted by deep surveys. One promising candidate that has been identified as part of this survey, Lupus-TR-3b, appears to be a Hot Jupiter with radius $0.9R_J$ and mass $0.8M_J$ based on its light curve and follow-up spectroscopy performed at the Magellan~II telescope. However, high resolution imaging data obtained with the Magellan~I telescope, combined with image deconvolution analysis, has revealed that the host ``star'' in fact consists of up to seven separate stars that are unresolved in the SuperLupus photometry \citep{2009IAUS..253...55S}. While four of these stars are too faint to be responsible for the observed transit signal, it is unclear which of the remaining three stars is undergoing eclipse. Therefore, while unlikely, Lupus-TR-3b can still not be completely ruled out as being a blended EcB rather than a transiting planet. At the very least, if Lupus-TR-3b is a planet, it will have different properties to those listed above, which were derived under the assumption that the signal was caused by a planet transiting a single star.   

These two examples illustrate the particular challenge posed to transit surveys by blending, both in terms of the potential for false detections to be made and the effect that it can have on estimated planetary properties. The \emph{a posteriori} processes used to identify blended neighbors, such as spectral line asymmetry analysis and iterative image deconvolution, are typically time-consuming and resource-intensive tasks, albeit important ones. Considering the problem in the forward direction, starting with well-understood statistics of stellar crowding and property distributions of multiple stars and Hot Jupiters provides another perspective on the problem, and is the approach taken in this work.

\section{Methodology}  \label{sec3}

\subsection{The synthetic surveys}

We investigate the rates of astrophysical false detections (i.e.\ EcBs that mimic a Hot Jupiter signal) and genuine Hot Jupiter detections separately for a synthetic deep survey, based on the SuperLupus survey \citep{2009IAUS..253..333B,2009AJ....137.4368B}, and a synthetic shallow survey, loosely based on surveys such as HAT-Net~\citep{2007ApJ...656..552B} and TrES~\citep{2004ApJ...613L.153A}. The important properties of each survey are listed in Table~\ref{tab:survs}. For simplicity, and ease of comparison, the field of views~(FOVs) of both synthetic surveys are taken to be centered on the same Galactic coordinates as the SuperLupus survey $l=331.5^\circ$, $b=11^\circ$. The magnitude range of the deep survey is taken to be $15<V<19$, the same as SuperLupus, while for the shallow survey a magnitude range of $10<V<13$ is used. Using these parameters, expected star counts are binned according to luminosity class, spectral type, and apparent~$V$ magnitude for each synthetic survey using the Besan\c{c}on model of the Milky Way Galaxy \citep{2003A&A...409..523R}. Main sequence stars are found to comprise approximately $75$\% of all stars in our synthetic deep survey, and about $20$\% in our synthetic shallow survey. 

\subsection{Estimating detection rates}

We estimate the rate of false detections~$\Omega_{FD}$ due to EcBs by integrating a probability density function~$\Phi_{FD}$ of the form:
\begin{eqnarray}
  \frac{d^6 \, \Omega_{FD}}{d\Psi\cdot dV_p\cdot dP\cdot dq\cdot d(\cos i)\cdot df_B} &=& \Phi_{FD}(\Psi,V_p,P,q,\cos i,f_B)   \label{eq:omfd}
\end{eqnarray}
where $\Psi$~is the spectral type of the primary, $V_p$~is the apparent magnitude of the primary, $P$~is the orbital period, $q$~is the secondary-to-primary mass ratio, $i$~is the orbital inclination, and~$f_B$ is the total blended flux. These properties determine the observed primary transit depth~$\delta_p$, observed secondary transit depth~$\delta_s$, and transit duration~$\tau$ (see \S\ref{sec:philc}). Hence in effect, the integrated density function~$\Phi_{FD}$ gives the probability that any given star monitored will be an EcB consisting of two dwarf components with light curve properties $\{\delta_p,\delta_s,\tau\}$ that satisfy our criteria for a false detection, which are presented in \S{\ref{sec:crit}. A detailed description of the integration process is given in  \S{\ref{sec:integrating}}. 

The rate of Hot Jupiter detections~$\Omega_{HJ}$ is estimated in exactly the same way, using a probability density function of the form: 
\begin{eqnarray}
  \frac{d^6 \, \Omega_{HJ}}{d\Psi \cdot dV_p\cdot dP\cdot dR\cdot d(\cos i)\cdot df_B} &=& \Phi_{HJ}(\Psi,V_p,P,R,\cos i,f_B), \ \  \label{eq:omhj}
\end{eqnarray}
where the binary mass ratio~$q$ is replaced with the planetary radius~$R$.

\subsection{EcB and Planet-Star configurations considered}  \label{sec:configs}

Ground-based transit surveys are overwhelmingly sensitive to transiting planets with periods $P\lesssim 10$~days, due to observational window effects as well as the low geometric probability of a transit occurring for longer periods. Therefore, we only consider EcBs and Hot Jupiters with periods in this range. This allows us to ignore EcBs with a giant component, as detached giant binaries do not exist for periods $P<10$~days \citep{2004A&A...421..241S}. We also only consider Hot Jupiters transiting main sequence stars, since the photometric signal caused by a Hot Jupiter transiting a giant star is too faint to be detected \citep{2003ApJ...585.1056G}.   

We divide both false detections and Hot Jupiter detections into two main types: blended and unblended. Blended detections are those for which the eclipsing system is accompanied by one or more unresolved neighbor stars. The resolution limit~$r_c$ is taken to be approximately equal to the size of the PSF or the pixel scale, whichever is largest. For the deep survey we use a value of $1.5''$, which is roughly equal to the spatial resolution achieved by the SuperLupus survey, and for the shallow survey we use a value of $20''$, which is typical of the PSFs in actual  shallow surveys \citep[eg.][]{2004PASP..116..266B}. Blending has the effect of reducing the measured transit depths~$\delta_m$, according to:
\begin{eqnarray}
  \delta_m &=& \frac{f_u}{f_u+f_B}\delta_u  \label{eq:deltam}
\end{eqnarray}
where $f_u$~is the flux of the unblended EcB, $f_B$~is the total flux of the blended neighbors, and $\delta_u$ is the transit depth that would be measured if the EcB was unblended. 

With regard to tertiary companions of EcBs, in reality some would fall within the confusion radius $r_c$ and some would fall outside. However, we chose not to estimate the true fractions of resolved and unresolved tertiary components, and instead perform the calculations separately under two limiting assumptions: (i) all tertiary companions are resolved, i.e.~fall outside the confusion radius~$r_c$ of the survey photometry; and (ii) all tertiary companions are unresolved. In the latter case, the tertiary companion is treated as an additional blended flux in Eq.~\ref{eq:deltam}.  Higher order companions are not accounted for.

We do not take stellar multiplicity into account when calculating Hot Jupiter detection rates. That is, we treat every star as single and ignore possible complications associated with detecting Hot Jupiter transits in multiple systems such as blending by binary companions. \cite{2008ApJ...686.1302B} also made this simplification, and estimated that it could result in an overestimation of planet detection rates by at most $30\%$. However, we do consider blended cases of Hot Jupiter detections (i.e. planet $+$ host star $+$ physically unassociated stars), in addition to unblended Hot Jupiter detections.

\subsection{Criteria for a detection} \label{sec:crit}

We define an astrophysical false positive as an EcB that meets all of the following criteria:
\begin{enumerate}
  \item Both stellar components are dwarfs with masses greater than $0.1 M_{\Sun}$, which approximately corresponds to the minimum hydrogen burning mass \citep{2000ARA&A..38..337C}. We do not include transits by brown dwarf companions in our false detection counts, since brown dwarfs orbiting within 5AU of main sequence stars are known to be extremely rare (the ``brown dwarf desert'') \citep{2000PASP..112..137M}.
  \item The observed primary transit depth~is $\delta_p<0.05$ and has S/N$>\gamma$, where $\gamma$~is some minimum detection threshold. While it is possible for a Hot Jupiter transiting a late M dwarf to cause a fractional flux change greater than $0.05$, such cases are rare, and generally disregarded by survey teams due to the high probability that they are caused by an EcB.
  \item The observed secondary transit depth~$\delta_s$  either has S/N$<\gamma$ or $\delta_p-\delta_s < \epsilon$, where $\epsilon$ is the measurement noise, i.e.\ it is either too shallow to be detected or it is indistinguishable from the primary eclipse.
  \item The transit duration $\tau$ is less than $0.25$\,d, which approximately corresponds to the time taken for a Hot Jupiter with a period less than 10\,d to undergo a central transit of the disk of an F dwarf. Hot Jupiter transits of dwarfs earlier than F, which have larger radii and hence longer transit durations, will generally be too shallow for detection by ground-based surveys \citep{2003ApJ...585.1056G}. EcBs with periods $P<10\,$d, on the other hand, are much more likely to have transit durations $\tau>0.25\,$d due to the large relative disk sizes of the eclipsing bodies.
  \item The effective magnitude~$V_{eff}$ of the EcB combined with any blended neighbors is within the magnitude limits of the survey.
\end{enumerate}

With the exception of the mass constraint on the secondary, our criteria for a Hot Jupiter detection are exactly the same as those listed above for an EcB false detection. Of course, the secondary eclipse that occurs when the planet passes behind the star does not produce an observable signal, so for a transiting Hot Jupiter condition (3) will always be satisfied. We also note that Hot Jupiter transits with primary transit depths $\delta_p>0.05$ or transit durations $\tau > 0.25$\,d are possible, but since such signals are far more likely to be caused by EcBs, survey teams generally discard such candidates if they arise.

\subsection{The extended magnitude range}  \label{sec:embf}

In addition to the ``formal'' survey magnitude range, which we denote by~$V_b<V<V_f$ (i.e.\ $10<V<13$ for our synthetic shallow survey, and $15<V<19$ for our synthetic deep survey), we define an extended magnitude range~$V_b<V<V_e$ for each survey, where $V_f<V_e$. This is to allow for the possibility that blended stars with apparent magnitudes fainter than the formal survey cut-off~$V_f$ can contribute to the observed photometric signal. The faint limit~$V_e$ of the extended magnitude range is taken to correspond to the flux~$f_e \sim \sigma_m$, where $\sigma_m$ is the lowest root-mean square~(rms) scatter~$\sigma$ in the photometry achieved over the formal magnitude range. That is, as long as a star has flux at least at the level of the noise, we account for its contribution in the total measured flux.

To obtain analytic expressions for the fractional rms~$\sigma$ as a function of apparent magnitude~$V$, we fit parabolic curves of the form:
\begin{eqnarray}
  \log_{10}\sigma &=& aV^2+bV+c   \label{eq:rms}
\end{eqnarray}
by eye to the scatter in light curves as a function of $V$ in the SuperLupus survey~\citep[Fig.~2,][]{2009AJ....137.4368B} and the TrES survey~\citep[Fig.~2.2,][]{odonovanphdthesis}, corresponding to our synthetic deep and shallow surveys, respectively. Values for $a$, $b$, and $c$ for each survey are listed in Table~\ref{tab:survs}. Using Eq.~\ref{eq:rms}, we find $V_e=22$ for the deep survey and $V_e=16$ for the shallow survey.

In order for any star with magnitude $V_f<V<V_e$ to contribute to a photometric measurement, however, it must be blended with at least one other star that has magnitude $V_b<V<V_f$, i.e.\ within the formal survey magnitude limits. To simplify the analysis, we ignore those cases in which two or more unresolved stars each have a brightness fainter than the faint limit of the survey, but a total brightness greater than the faint limit. Such ``borderline'' cases will be relatively rare and should not have a significant effect on the final calculated result. 

Supposing we have a FOV containing a total of $N$~stars, the probability~$p_n$ that any one of those stars will be blended with precisely~$n$ other stars is given by
\begin{eqnarray}
  p_n &=& \frac{(N-1)!}{n!\,(N-n-1)!\,M^n}\left( 1-\frac{1}{M} \right)^{N-n-1}   \label{eq:pn}
\end{eqnarray}
where $M=A_{FOV}/\pi r_c^2$ is the number of resolution elements in a FOV of area~$A_{FOV}$, given a confusion radius~$r_c$. 

By letting $N$~in Eq.~\ref{eq:pn} be equal to the total number of stars~$\alpha$ in the FOV (both resolved and unresolved) with magnitude $V_b<V<V_f$, we obtain the fraction $\tilde{p}_n$ of stars that are blended with exactly $n$~other stars with magnitudes in the formal magnitude range. This means that if $\beta$ is the number of stars with magnitude $V_f<V<V_e$ between the faint limit $V_f$ of the formal magnitude range and the faint limit $V_e$ of the extended magnitude range, then only $(1-\tilde{p}_0)\beta$ of them will be able to contribute to the photometric measurements, i.e.\ those that are blended with at least one other star within magnitude in the formal range. The effective total star count~$N_\Sigma$ over the extended magnitude range is therefore:
\begin{eqnarray}
  N_\Sigma &=& \alpha + (1-\tilde{p}_0)\beta  \label{eq:nsig}
\end{eqnarray}  
and the number of resolved objects~$N_R$ is given by:
\begin{eqnarray}
  N_R &=& \alpha \, \sum_n{\frac{\tilde{p}_n}{n+1}}  \label{eq:nr}
\end{eqnarray}
In \S\ref{sec:rates}, we use Eq.~\ref{eq:nr} to quote our detection rates as the number of detections per $N_R=10,000$ resolved stars.

 Values for the crowding probability~$p_n$ in each synthetic survey computed using Eq.~\ref{eq:pn} and~\ref{eq:nsig} are listed in Table~\ref{tab:nblends}. The low blending probabilities~$p_n$ for $n \geq 1$ we obtain for the synthetic deep survey could mean that the case of Lupus-TR-3b, which consists of up to seven blended stars (\S\ref{sec2}), is an unlikely single event. On the other hand, of the seven blended stars, at least four have magnitudes $V>22$, making them too faint to be accounted for in our analysis which imposes a strict cut-off at $V=22$. Two of the remaining blends may also be a wide physical binary pair \citep{2009IAUS..253...55S}. It is also possible that the star counts we obtain from the Besan\c{c}on model are underestimated. For instance, the actual interstellar extinction may be lower than the value of 0.7\,mag\,kpc${}^{-1}$ that we use for the Besan\c{c}on model input.

\subsection{Integrating the probability density function}  \label{sec:integrating}

To estimate detection rates, we integrate the probability density functions given by Eq.~\ref{eq:omfd} and \ref{eq:omhj} over appropriate property ranges (see below). Due to the independence of parameters in these relations, the probability density functions $\Phi_{FD}$ and $\Phi_{HJ}$ can be broken into a number of component probabilities. For the false detections:
\begin{eqnarray}
  \Phi_{FD}(\Psi,V_p,P,q,\cos i,f_B) &=& \Phi_{d}(\Psi,V_p) \cdot \Phi_{b}(P,q) \cdot \Phi_i(\cos i) \cdot \Phi_f(f_B)  \nonumber \\
  && \hspace{1.0cm} \times \ \ \Phi_w(P) \cdot \Phi_{lc}(\Psi,V_p,P,q,\cos i,f_B)  \label{eq:phifd}
\end{eqnarray} 
where $\Phi_d(\Psi,V_p)$ is the probability that the star will be main sequence with spectral type~$\Psi$ and apparent magnitude~$V_p$; $\Phi_b(P,q)$ is the probability that a main sequence star will be the primary component of a multiple system with period~$P$ and mass ratio~$q$; $\Phi_i(\cos i)$ is the probability that the orbital inclination is~$i$; $\Phi_f(f_B)$~is the probability that the total flux due to blended companions is~$f_B$; $\Phi_w(P)$ is the probability that at least three transits will be observed given the observing window function; and $\Phi_{lc}(\Psi,V_p,P,q,\cos i,f_B)$ is the probability that the transit signal of a binary consisting of two dwarfs with properties~$\{\Psi,V_p,P,q,\cos i,f_B \}$ resembles that of a transiting Hot Jupiter. The assumption of independence between the spectral type~$\Psi$ and the other properties of the binary, namely, the period~$P$ and mass ratio~$q$, is supported by empirical studies~\citep{2003A&A...397..159H}. 

Similarly, for the planet detections, we can divide the probability density function of Eq.~\ref{eq:omhj} into: 
\begin{eqnarray}
  \Phi_{HJ}(\Psi,V_p,P,R,\cos i,f_B) &=& \Phi_{d}(\Psi,V_p) \cdot \Phi_{b}(P,R) \cdot \Phi_i(\cos i) \cdot \Phi_f(f_B)  \nonumber \\
  && \hspace{1.0cm} \times \ \ \Phi_w(P) \cdot \Phi_{lc}(\Psi,V_p,P,R,\cos i,f_B)  \label{eq:phihj}
\end{eqnarray}
where $\Phi_{b}(P,R)$ refers to the probability that a main sequence star will host a Hot Jupiter companion with orbital period~$P$ and radius~$R$, and all other terms are the same as for the false detection case given by Eq.~\ref{eq:phifd}. Again, we assume independence between the spectral type of the primary~$\Psi$ and the companion properties, which in this case are the planet radius~$R$ and orbital period~$P$. While correlations are beginning to emerge between system parameters as more transiting planet systems are discovered, such as between stellar mass and planet mass \citep{2007ApJ...670..833J}, and planet mass and period \citep{2008ApJ...677.1324T}, we note that the observed scatter is still large\footnote{http://exoplanet.eu/}. Therefore, while our assumption of independence between the system parameters mentioned above is a simplification, it is reasonable for the purpose of this study and is unlikely to have a substantial effect on our final results.

For the EcB false detections, integration is performed numerically over the following ranges:
\begin{itemize}
  \item All spectral types $\Psi \in \left[\mbox{O}0,\mbox{M}9  \right]$, with a step size $d\Psi$ of five spectral sub-types.
  \item The extended magnitude range $V_p \in \left[ 15,22 \right]$ for the deep survey and $V_p \in \left[ 10,16 \right]$ for the shallow survey, with a step size $dV_p=0.5$ mag.
  \item Periods $P \in \left[ 1,10 \right]$, with a step size $dP=0.5$ days.
  \item Mass ratios $q \in \left[ 0,1 \right]$ with a step size $dq=0.1$.
  \item Orbital inclinations $\cos i \in \left[ 0,1 \right]$ with a step size $d(\cos i) = 0.01$.
  \item For the blended cases, fluxes $f_B \in \left[ f_f,f_b \right]$ where $f_f$~is the flux corresponding to the faint magnitude limit of the extended magnitude range and $f_b$~is the flux corresponding to the bright magnitude limit. A step size $df_B$ corresponding to half a magnitude is used. For the unblended cases, $f_B$ is not integrated over, i.e. $\Phi_f(0)=1$ in Eq.~\ref{eq:phifd} and~\ref{eq:phihj}. 
\end{itemize}

For the Hot Jupiter detections, spectral types~$\Psi \in [\mbox{F}0,\mbox{M}9]$ are integrated over, since dwarfs of spectral type earlier than about $\mbox{F}0$ are too large for a Hot Jupiter transit to be detectable \citep{2003ApJ...585.1056G}. We integrate over Hot Jupiter radii~$R\in [0.8,1.6]R_J$ with a step size $dR=0.1R_J$. We choose this range of radii because it encompasses the vast majority of Hot Jupiters discovered via transits at the time of writing\footnote{http://exoplanet.eu/}. All other integration ranges are the same as for the EcB false detections listed above.

In the following subsections, we describe the observational data and assumptions that are used to define the individual components of the probability densities in Eq.~\ref{eq:phifd} and~\ref{eq:phihj}.

\subsubsection{Stellar property distribution $\Phi_d(\Psi,V_p)$}

We use star counts from the Besan\c{c}on model \citep{2003A&A...409..523R} to calculate the probability $\Phi_d(\Psi,V_p)\,d\Psi\,dV_p$ that a star will be main sequence with spectral type $\Psi$ and apparent magnitude~$V_p$:
\begin{eqnarray}
  \Phi_d(\Psi,V_p) &=& \frac{N(\Psi,V_p)}{N_{\Sigma}} 
\end{eqnarray}
where $N(\Psi,V_p)$ is the number of main sequence stars with spectral type~$\Psi$ and  apparent magnitude~$V_p$, and $N_{\Sigma}$ is the number of stars of all luminosity classes, spectral types, and magnitudes in the extended magnitude range given by Eq.~\ref{eq:nsig}.

\subsubsection{Companion property distributions $\Phi_b(P,q)$ and $\Phi_b(P,R)$}

We assume that the period~$P$ of a stellar binary is independent of the mass ratio~$q$, i.e.
\begin{eqnarray}
  \Phi_b(P,q) &=& \Phi_P(P)\cdot\Phi_q(q)  \label{eq:phipq}
\end{eqnarray}
where $\Phi_P(P)$ is the probability that a main sequence star will be the primary of a binary system of period~$P$, and $\Phi_q(q)$ is the probability that a main sequence binary system will have mass ratio~$q$. The distributions for $\Phi(P)$ and $\Phi_q(q)$ are taken from the empirical study of spectroscopic binaries by \cite{2003A&A...397..159H}, who found no evidence for the period being dependent on the mass ratio over the range $0<P<10$\,d. This is consistent with our assumption of independence between these two properties (Eq.~\ref{eq:phipq}). When performing calculations for the limiting case of all tertiary companions being unresolved~(see \S\ref{sec:configs}), we take the fraction of binary systems with tertiary components as a function binary period from \cite{2006A&A...450..681T}. 

Similarly, we assume that the period~$P$ and radius~$R$ of a Hot Jupiter are independent, i.e.
\begin{eqnarray}
  \Phi_b(P,R) &=& \Phi_P(P)\cdot\Phi_R(R) 
\end{eqnarray}
where $\Phi_P(P)$ is the probability that an F0-M9 main sequence star will host a Hot Jupiter of period~$P$, and $\Phi_R(R)$ is the probability that a Hot Jupiter will have radius~$R$. The distribution for $\Phi_P(P)$ is obtained from the empirical estimates of \cite{2007A&A...475..729F}, and for simplicity we assume a uniform distribution for~$\Phi_R(R)$ in the range of radii we consider here, namely 0.8-1.6\,$R_J$. We chose to take the period distribution $\Phi_P(P)$ of \cite{2007A&A...475..729F}, which was derived from the results of the OGLE transit survey, as opposed to analogous distributions obtained from radial velocity surveys \citep[eg.][]{2008PASP..120..531C} due to the inherent metallicity bias in the latter. See \S{2.2} of \cite{2008ApJ...686.1302B} for a fuller discussion of this issue.

\subsubsection{Inclination distribution $\Phi_i(\cos i)$}

Orbits are assumed to be orientated randomly in three-dimensional space, giving $\Phi_i(\cos i)=1$.

\subsubsection{Blending distribution $\Phi_f(f_B)$}  \label{sec:phifb}

The probability~$\Phi_f(f_B)$ that a star is blended with one or more other stars with a total blended flux~$f_B$ is
\begin{eqnarray}
  \Phi_f(f_B) &=& \sum_n{\tilde{\Phi}_f(n,f_B)}   \label{eq:phifb1}
\end{eqnarray}
where $\tilde{\Phi}_f(n,f_B)$ is the probability that the blend consists of exactly $n$~other stars with total flux $f_B$, and is given by
\begin{eqnarray}
   \tilde{\Phi}_f(n,f_B) &=& \sum_{\varphi}{\left[ \prod_{i=1}^n{N_i} \right] }\frac{1}{M^n}\left( 1-\frac{1}{M} \right)^{N_{\Sigma}-n-1}  \label{eq:phifb2}
\end{eqnarray} 
where $\varphi$ denotes all combinations of fluxes $f_1, f_2, \ldots, f_n$ that satisfy
\begin{eqnarray}
  f_B &=& \sum_{i=1}^n{f_i} 
\end{eqnarray}
$N_i$ is the number of stars in the field with flux $f_i$, $M$~is the same as in Eq.~\ref{eq:pn}, and $N_\Sigma$ is the effective total number of stars in the FOV as given by Eq.~\ref{eq:nsig}. Values for the blended flux probabilities $\Phi_f(f_B)$ computed using Eq.~\ref{eq:phifb1} and~\ref{eq:phifb2} are plotted for each synthetic survey in Fig.~\ref{fig:fblends}.

\subsubsection{Window function $\Phi_w(P)$} \label{subsec:window}

For both surveys, we use the window function~$\Phi_w(P)$ provided in Fig.~2 of \cite{2009IAUS..253..333B}. This function is an approximation of the real window function for the SuperLupus survey, and is calculated assuming 100 contiguous observing nights with weather statistics that match actual weather logs from the Siding Spring Observatory site.

\subsubsection{Light curve detection probabilities $\Phi_{lc}(\Psi,V_p,P,q,\cos i,f_B)$ and $\Phi_{lc}(\Psi,V_p,P,R,\cos i,f_B)$} \label{sec:philc}

Assuming that the orbit is circular, and given the properties $\{\Psi,V_p,P,q,\cos i,f_B\}$, we can determine the observed transit depths, $\delta_p$ and~$\delta_s$, and the transit duration~$\tau$ as follows. First, we use the spectral type~$\Psi$ of the primary to find its mass~$M_p$, radius~$R_p$, and luminosity~$L_p$  using a fit to the relations for main sequence stars tabulated by~\cite{2000asqu.book.....C}. The luminosity combined with the apparent magnitude~$V_p$ then provides the distance~$d$ to the system. The primary mass and the mass ratio~$q$ yield the secondary mass~$M_s$, and hence the secondary radius~$R_s$ and luminosity~$L_s$. The distance~$d$ and secondary luminosity~$L_s$ is next used to calculate the flux~$f_s$ from the secondary. The masses, radii, and period~$P$ provide the orbital semi-major axis~$a$ via Kepler's Third Law.

With these properties, we have enough information to compute the observed primary transit depths, $\delta_p$ and $\delta_s$, and the transit duration~$\tau$ using standard relations such as those set out in \cite{1999poss.conf..189S}. We assume a quadratic limb darkening law with solar coefficients taken from \cite{2000asqu.book.....C}, regardless of spectral type. 

The S/N of the transit signature in each light curve is then calculated for each of $\delta_p$ and $\delta_s$ via:
\begin{eqnarray}
  \mbox{S/N} &=& \frac{\delta_i}{\sigma}\sqrt{\mu}   \label{eq:sn}
\end{eqnarray}
where~$\sigma$ is the measured out-of-transit photometric rms scatter given by Eq.~\ref{eq:rms}, and~$\mu$ is the number of in-transit data points, which we take to be:
\begin{eqnarray}
  \mu &=& \frac{\eta h \upsilon}{\omega}\frac{\tau}{P} 
\end{eqnarray}
where $\eta$~is the number of observing nights, $h$~is the observing time per night, $\upsilon$~is the fraction of usable nights, and $\omega$~is the measurement cadence. For simplicity, and ease of comparison, we use $\eta=100$ nights, $h=8$ hours, $\upsilon=0.67$, and $\omega=6$ minutes for both synthetic surveys, which is comparable with the actual experience of the SuperLupus survey \citep{2009IAUS..253..333B,2009AJ....137.4368B}. 

Equipped with the transit depths $\delta_p$ and~$\delta_s$, the associated S/N ratios, the transit measurement noise~$\epsilon\equiv \sigma / \sqrt{\mu}$ and the transit duration, we are then able to determine if the EcB configuration satisfies the criteria for a detection given in \S{\ref{sec:crit}}. The probability is then simply:
\begin{eqnarray}
  \Phi_{lc}(\Psi,V_p,P,q,\cos i,f_B)=\left\{
  \begin{array}{cl}
  1 & \ \ \ \textnormal{if the detection criteria are met} \\
  & \\
  0 & \ \ \ \textnormal{otherwise}
  \end{array}
  \right.\label{eq:thetap}  
\end{eqnarray}

The process for the Hot Jupiter detections is directly analogous to that outlined above for false detections. The only difference is that instead of a mass ratio~$q$ being used to obtain the secondary mass, the Hot Jupiter mass~$M$ is obtained from the radius~$R$ by assuming all Hot Jupiters have the same density~$\rho_J$ as Jupiter, i.e. $M=4\pi R^3 \rho_J/3$. This is a simplification, considering the observed range in mean densities of the transiting planets discovered so far, from abnormally-inflated planets such as TrES-4 with only $\sim$0.1 times the mean density of Jupiter \citep{2007ApJ...667L.195M}, to objects such as XO-3b \citep{2008ApJ...677..657J} and WASP-18b \citep{2009Natur.460.1098H} with mean densities $\sim$10 times that of Jupiter. However, this simplifying assumption will only have a minor effect on the final results. This is because we only use the mass of the planet in the calculation of the orbital semi-major axis~$a$, which is then used with the orbital inclination $\cos i$ to calculate the transit latitude $a \cos i$. Since the semi-major axis is $a\sim \left( M_p+M_s \right)^{1/3}$ for a given orbital period~$P$ according to Kepler's Third Law, and $M_s << M_p$ in the case of a star-planet system, the uncertainty in the mass of the planet will only have a negligible effect on the calculated semi-major axis and hence on the calculated transit latitude. 

It must be noted that Eq.~\ref{eq:sn} assumes Poisson statistics for the noise, i.e.\ that the noise decreases according to $\sim 1/\sqrt{\mu}$ (white noise). In practice, however, effects such as varying airmass, weather conditions, moon phase, and telescope tracking cause systematic trends in the measured light curves that can be correlated over the timescale of a planetary transit (red noise) \citep{2006MNRAS.373..231P}. This means that a measured light curve will in fact have a lower S/N than predicted by Eq.~\ref{eq:sn}. Rather than attempting to model the white noise and red noise components individually, as was done by \cite{2008ApJ...686.1302B}, we simply require that the underestimated S/N calculated using Eq.~\ref{eq:sn} exceeds a higher cut-off than a survey team would typically require. Specifically, we calculate separate detection rates for S/N cut-offs of 10, 20, and 30. By varying the S/N cut-off, we are also able to investigate the sensitivity of the yield to this component of the detection criteria.

\section{Results}

\subsection{Predicted rates}  \label{sec:rates}

Calculated detection rates for the synthetic deep and shallow surveys obtained using S/N detection cut-offs of 10, 20, and 30 are reported in Tables~\ref{tab:tcrrates} and~\ref{tab:tcurates}, assuming separately that all tertiary components in multiple star systems are resolved and unresolved, respectively (\S\ref{sec:configs}). Values are expressed as the number of detections per $10,000$ resolved stars monitored with sufficient photometric precision to detect a Hot Jupiter transit. Recall that the number of resolved stars~$N_R$ in the survey field is given by Eq.~\ref{eq:nr}. Values without parentheses are those that have been calculated with a window function~(\S\ref{subsec:window}), while those values with parentheses have been calculated ignoring window effects. In both synthetic surveys, the inclusion of our window function does not significantly reduce the predicted number of detections. This reflects the fact that the window function we used~(\S\ref{subsec:window}) gives almost complete recovery for signals with periods $\lesssim5$~days.

The results in Tables~\ref{tab:tcrrates} and~\ref{tab:tcurates} show that the total false detection rates are hardly changed if we assume that all tertiary components are resolved or unresolved in the survey photometry. This is not surprising, considering that in our calculations, an unresolved tertiary component only has the effect of increasing the overall flux by an amount less than or equal to that of the primary (since we assume a mass ratio $q_3 \leq 1$, \S{\ref{sec:configs}). Therefore, some of the EcB transits that are too deep to be mistaken for a planetary signal if the diluting effect of the tertiary component is not accounted for will become shallow enough to cause a false detection when the additional flux is included (Eq.~\ref{eq:deltam}). On the other hand, some of the EcB transits that are only just deep enough to be detected when they are not blended with an unresolved tertiary component will instead fall below the noise when the additional flux of an unresolved tertiary component is included in the transit depth calculation. The fact that our overall false detection rates show hardly any change if we assume all tertiary components are resolved or unresolved indicates that these competing effects are roughly equal in magnitude.

The results in Table~\ref{tab:tcurates} suggest that hierarchical triple systems are likely to constitute a substantial fraction of false detections in transit surveys close to the galactic plane. For the limiting case that all tertiary components are unresolved, in both synthetic surveys we find that hierarchical triples cause $\sim$5 times more false detections than EcBs without tertiary components. This result can be explained by the fact that short period binaries are highly likely to have tertiary components \citep{2006A&A...450..681T}, and they also have a relatively high geometric probability of transiting \citep[see Eq.~2 of ][]{1999poss.conf..189S}.

Overall, we find $\sim$2-4 false detections and $\sim$1-3 Hot Jupiter detections per $10,000$ stars monitored in the deep synthetic survey, depending on the S/N cut-off used to define a detection (\S\ref{sec:philc}). False detections are divided roughly evenly between blended and unblended detections, whereas only about $20\%$ of planet detections are blended. 

These results compare well with the SuperLupus survey in which a total of 13 planetary candidates have been identified from a sample of $\sim$25,000 stars monitored~(Bayliss 2009, private communication). Of these, however, at least five do not meet the all of the detection criteria used in the present study. They either have transit durations longer than $0.25$~days or secondary eclipses that were detected once greater phase coverage was obtained for the orbit. This leaves a total of eight candidates, six of which have not yet been followed up. Of the two candidates that have been followed up, one  is most likely a transiting Hot Jupiter (Lupus-TR-3b, \S\ref{sec2}) and the other has been confirmed as a false positive. 

In the synthetic shallow survey, we find $\sim$3-4 false detections, similar to the synthetic deep survey, but only $\sim$0.1-0.6 Hot Jupiter detections per $10,000$ stars monitored. In other words, one planet is detected for every $\sim$15,000-100,000 stars monitored, and for every planet detected we obtain $\sim$5-25 false detections, depending on the S/N cut-off used in the calculations. Approximately $90\%$ of the false detections in the shallow survey are caused by blended EcBs, whereas blended and unblended planet detections occur in roughly equal amounts.

These results seem broadly consistent with the commonly-reported experience of shallow surveys that astrophysical false positives outnumber planet detections by an order of magnitude. For instance, as part of the TrES survey, \cite{odonovanphdthesis} used the 10cm Sleuth telescope to monitor $\sim$150,000 stars across 19 fields with photometric precision $\lesssim 2\%$. From this, 67 candidates were identified but only 4 were confirmed as genuine planet detections. By comparison, our results predict for every $150,000$ stars monitored there will be $\sim$2-12 Hot Jupiter detections and $\sim$40-70 false detections, depending on the S/N cut-off used and whether or not a window function is included. The agreement to within a factor of $\sim$2 is reasonable, considering the idealized nature of our detection criteria, our assumption that full orbital phase coverage is obtained, and the fact that our calculations were made for a single field location whereas Sleuth monitored several fields spread over a range of galactic latitudes.

Another point worth noting about the results for the synthetic shallow survey listed in Tables~\ref{tab:tcrrates} and~\ref{tab:tcurates} is that the rate of false detections caused by unblended configurations remains constant within the numerical precision as the S/N cut-off is increased from 10 to 30. This is due to the fact that as the S/N cut-off is increased, less primary transits \emph{and} less secondary transits are detected above the noise. The former has the effect of decreasing the false detection rate, whereas the latter has the effect of increasing it. Since no net change is observed in the calculated detection rates, both opposing effects must be equal in magnitude for the unblended false detections in the shallow survey.

More specifically, the type of configuration that is likely to cause a false detection at low S/N cut-offs but not at higher S/N cut-offs are those for which: (i)~the primary transit signal is only just deep enough to be detected above the noise at the lower S/N cut-offs but not at the higher cut-offs; and (ii)~the secondary transit signal is either shallow enough to escape detection or is indistinguishable from the primary transit. Such configurations will lead to a reduction in false detections as the S/N cut-off is increased. At the same though, there will be configurations for which the primary transit is deep enough to be readily detectable (i.e.~S/N$\geq30$) but not so deep that it is  ruled out as a planetary candidate (i.e. $\delta_p<0.05$), while the secondary transit is only shallow enough to evade detection at the higher SN cut-offs. Such cases will increasingly contribute to the number of false detections as the S/N cut-off is increased. Both of the situations described here can arise from grazing configurations of an F or G star and a smaller star such as a late K or M dwarf (see \S{\ref{sec:dtrends}}), the difference being that the transit latitudes will be lower in the former case than they are in the latter.  

In the synthetic deep survey, however, the unblended false detection rate decreases as the S/N cut-off is increased from 10 to 30. The reason for this is the lower photometric precision attained in the deep survey compared to the shallow survey (Eq.~\ref{eq:rms} and Table~\ref{tab:survs}). As a result, there are relatively fewer configurations with a distinct secondary transit that is detectable at lower S/N cut-offs but not detectable at higher S/N cut-offs than in the shallow survey. Instead, most of the unblended false detections in the deep survey have primary transit depths that are only just detectable above the noise at the lower S/N cut-off, and secondary transits that are either too shallow to be detected or of a similar depth to the primary transit, as in the first case described above for the synthetic shallow survey. 

For both synthetic surveys, the rate of blended false detections decreases as the S/N cut-off is increased. Again, this is due to configurations with primary transits that are detectable at the low S/N cut-offs becoming undetectable as the cut-off is increased, and a secondary transit that is either too shallow to be detected above the noise or not discernible from the primary transit. This mainly occurs for cases in which significantly more flux comes from the blended stars than from the EcB itself (\S{\ref{sec:dtrends}}), resulting in shallower transit depths. Meanwhile, since planetary occultations only exhibit a primary transit signal, the rates of blended and unblended planet detections also decrease as the S/N cut-off is increased in both surveys.

We emphasize that our current study is intended to be illustrative: it should not be assumed that the results reported here for a FOV centered on galactic latitude $b=11^\circ$ also apply to surveys conducted further away from the galactic plane, where the stellar population statistics and crowding probabilities will be different. A proper investigation of how detection rates are affected by the galactic latitude of the survey field is left to a future analysis.

\subsection{Predicted detection trends} \label{sec:dtrends}

The rate of false detections and Hot Jupiter detections in both the synthetic shallow and deep surveys are shown in Fig.~\ref{fig:res12} as a function of the spectral type of the primary star being eclipsed. These distributions closely follow the star counts returned by the Besan\c{c}on model. However, in both surveys, the planet detections are skewed towards late-type dwarfs. This is due to the deeper, and hence more readily detectable, eclipses that result when Hot Jupiters transit late-type dwarfs which have comparatively small radii and low luminosities. For the false detections, we found that increasing the S/N cut-off between 10 and 30 biases the relative fraction of detections slightly towards earlier spectral types (F5 and G0). This occurs because the photometric rms scatter decreases towards brighter magnitudes (\S{\ref{sec:embf}}) where the fraction of earlier spectral types increases relative to later spectral types. However, this effect is quite small, with the relative fractions of detections changing by $\lesssim 5$\% for a given spectral type, and so we have only included the plotted results for the S/N$\geq20$ cut-off scenario in Fig.~\ref{fig:res12}.   

Fig.~\ref{fig:dres34} and~\ref{fig:sres34} show the rate of false detections and Hot Jupiter detections as a function of effective magnitude in the deep synthetic survey and shallow synthetic survey, respectively. Results are shown for the S/N$\geq10$ and S/N$\geq30$ cases to illustrate the effect of that varying the S/N cut-off has on the relative fractions of detections. In both synthetic surveys for the S/N$\geq10$ cut-off and S/N$\geq30$ cut-off, the rate of false detections and Hot Jupiter detections initially increases with increasing magnitude, simply reflecting the increasing numbers of stars at fainter magnitudes. A maximum is then reached, followed by a decline in detection rates over the faintest portion of the survey magnitude range. The decline in rates occurs due to the decrease in photometric precision at fainter magnitudes. For the same reason, Fig.~\ref{fig:dres34} and~\ref{fig:sres34} show that the peak in the relative fraction of detections is shifted towards successively brighter magnitudes as the S/N cut-off used to define a detection is increased from 10 to 30. An interesting feature of Fig.~\ref{fig:dres34} and~\ref{fig:sres34} is that the peak in the relative fraction of planet detection rates tends to occur at brighter magnitudes with higher photometric precisions than it does for false detections, especially as the S/N detection cut-off is increased. This is because the number of EcBs with $\delta_p\sim0.05$ is greater than for Hot Jupiter transits, with the latter tending to produce shallower eclipses that are harder to detect above the noise. 

Fig.~\ref{fig:res56} shows how the number of blended detections varies with the fraction of blended light in both the synthetic deep and shallow surveys. We found that increasing the S/N detection cut-off slightly shifted the relative fraction of false detections towards fainter blends (i.e.~blends contributing less of the overall flux) and the relative fraction of Hot Jupiter detections towards brighter blends. However, as in the case of relative detection rates versus primary spectral type (Fig.~\ref{fig:res12}), this effect is small ($\lesssim 5$\%), so we have only plotted the results for S/N$\geq20$ in Fig.~\ref{fig:res56}. These results show that around $35\%$ of the blended false detections in the deep survey are due to bright blends; namely, those blended false detections in which the EcB contributes $<30\%$ of the total flux. In the shallow survey, roughly $80\%$ are due to bright blends. For blended planet detections in both surveys we find that blends of up to about $50\%$ can occur, which has implications for planetary parameter estimates (see \S\ref{sec:disc}).

The EcB configurations responsible for unblended and blended false detections are illustrated in Fig.~\ref{fig:res7} and~\ref{fig:res8}, respectively. The primary purpose of these plots is to qualitatively illustrate the most common false detection configurations. Results are only shown for the case of S/N$\geq20$ since no appreciable change is observed as the S/N detection cut-off is increased from 10 to 30. We see that the majority of unblended false detections are caused by grazing ``twins'', which have indistinguishable primary and secondary eclipses. The remainder of cases are made up of: (i) grazing transits by stars with different radii; and (ii) low-latitude transits of a primary by a much smaller companion. For the blended false detections, there are two broad categories: (i) bright blends with a low-latitude EcB; and (ii) fainter blends with a grazing EcB.

\section{Discussion and Implications} \label{sec:disc}

The estimated false detection rates listed in Tables~\ref{tab:tcrrates} and~\ref{tab:tcurates} are similar for the deep and shallow synthetic surveys, but the estimated planet detection rates are $\sim$5 times higher for the deep survey than they are in the shallow survey. There are two main reasons for this: (i) $\sim$75\% of all stars in the deep survey FOV are dwarfs, whereas only $\sim$20\% are dwarfs in the shallow survey FOV, according to the Besan\c{c}on model star counts; and (ii) it is more difficult to detect a blended transiting Hot Jupiter than an unblended one, and the blend probability is higher in the shallow survey due to the larger confusion radius (Table~\ref{tab:nblends}). In practice, however, the comparatively low ratio of false detections to planet detections in the deep survey is at least partially offset by the inevitable difficulty of conducting suitable follow-up observations to verify or reject their faint candidates. 

Extrapolating from our synthetic surveys, $\sim$20\% of all false detections in shallow surveys and $\sim$65\% of all false detections in deep surveys close to the galactic plane will be due to EcBs that are either unblended or which contribute more than $30\%$ of the light in a blended configuration. For the bright targets of shallow surveys, such cases will generally display radial velocity variations that are readily detectable with $1$m-class telescopes. For the faint targets of deep surveys, however, acquiring follow-up spectroscopy with sufficient S/N will always be a challenge, but these configurations present the most accessible configurations to rule out.

We find that $\sim$65\% of all false detections in shallow surveys close to the galactic plane will be caused by bright blends with a physically unassociated giant. The equivalent fraction is $\sim$10\% in deep surveys close to the galactic plane, reflecting the lower frequency of giants over fainter magnitude ranges.  Typically, such cases are relatively easy to rule out, either through spectral typing or reference to existing stellar catalogs, at least in the case of the bright shallow survey targets. 

The other bright blend false detections, which constitute $\sim$15\% of all false detections in shallow surveys and $\sim25\%$ of all false detections in deep surveys close to the galactic plane, will be due to blending with physically unassociated dwarfs. It is this latter case of blended EcBs that pose the greatest challenge to follow-up observations. Multi-color photometry can often be useful to reveal a color-dependence in the eclipse depth. However, the task of unmasking such false positives would be considerably more difficult if the colors of the EcB components happen to be similar and the blended dwarf is a rapid rotator, in which case its broadened spectral lines may conceal those of the fainter EcB. 

\section{Conclusion}

We have investigated the rates of both false detections caused by EcBs and genuine Hot Jupiter detections in ground-based transit surveys from a bottom-up perspective, using empirically-determined property distributions of multiple star systems and Hot Jupiter systems as our input. In particular, we find that in both deep and shallow surveys: (i) significant numbers of false detections can be caused by faint EcBs blended with the light of one or more foreground dwarfs, a configuration which experience shows can produce particularly insidious impostors, and (ii) for genuine Hot Jupiter detections, up to $\sim$50\% of the system's light can be due to unassociated blends, which will affect the estimated planetary parameters if they are not accounted for. We therefore reiterate the conclusion of \cite{2009IAUS..253...55S}: namely, that it is important to identify or rule out blended neighbors for all candidates, including blended neighbors that are up to $\sim$5 magnitudes fainter than the primary. 

Lastly, we note that the space-based missions CoRoT \citep{2007AIPC..895..201B} and Kepler \citep{2003ASPC..294..427B} may be particularly susceptible to false detections or planet parameter misestimation caused by blending. Due to the high photometric precision of these instruments, blended neighbors \emph{up to 8 to 10 magnitudes fainter than the primary} can make non-negligible contributions to the measured light curves if the signal that is being sought is that of an Earth-sized object transiting a dwarf star. Furthermore, since these surveys target planets with masses down to about an Earth mass, and the monitored stars are quite faint ( $V \lesssim 15$-$16$), radial velocity follow-up will be of little use in confirming their low-mass candidates. Constraining the presence of faint blended neighbors will therefore be of particular importance if a terrestrial-sized planet is to be secured with confidence in these surveys.

\acknowledgements T.M.E. would like to thank The University of Sydney Department of Physics for hosting him during the final stages of the writing of this paper.

\bibliographystyle{apj}
\bibliography{mybib}

\begin{figure}
  \begin{center}
    \subfigure{
      \includegraphics[width=0.45\linewidth]{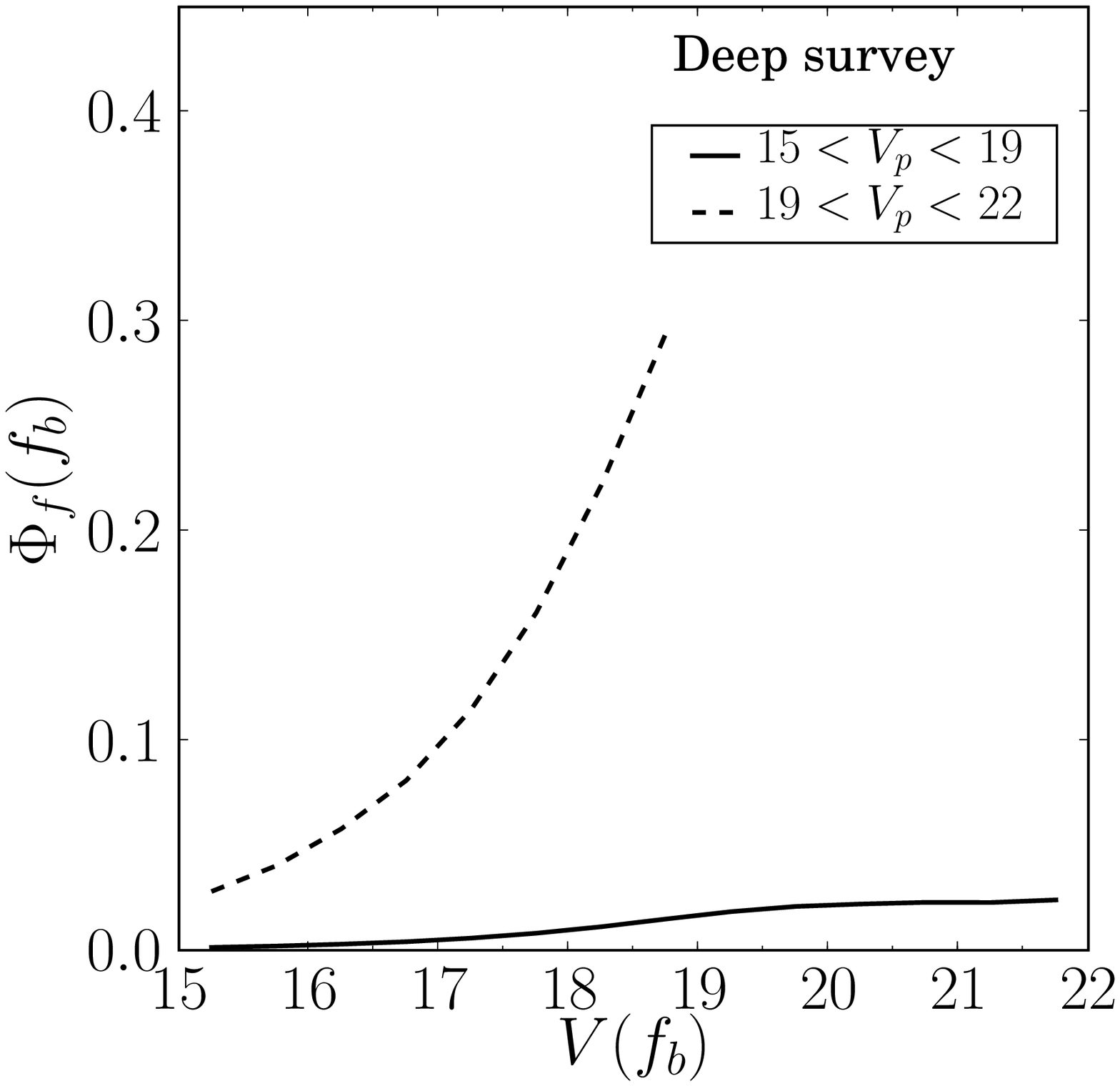}
      \label{}
      }
    \subfigure{
      \includegraphics[width=0.45\linewidth]{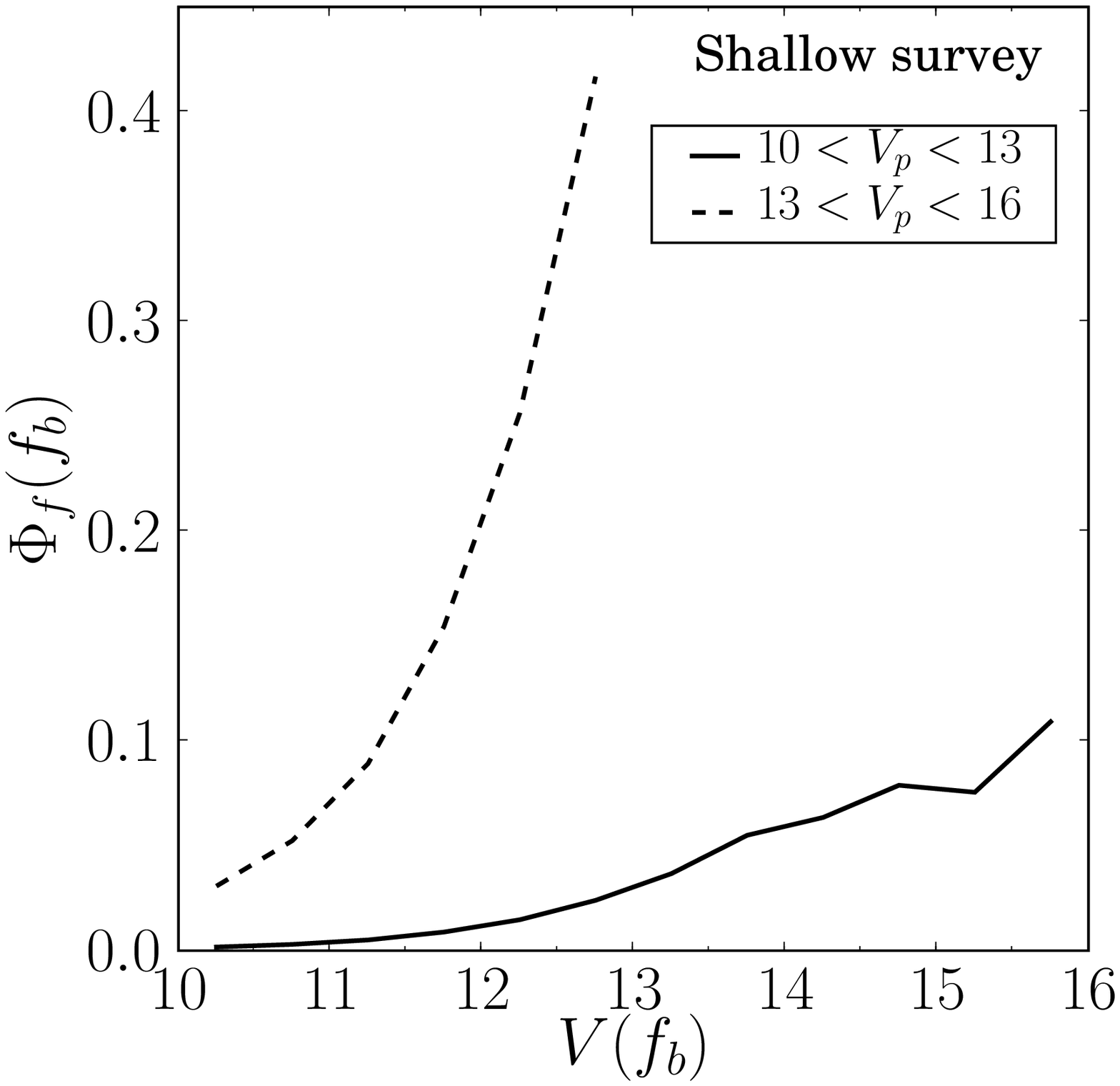}
      \label{} 
      }
    \caption{Curves for the deep (\emph{left}) and shallow (\emph{right}) synthetic surveys showing the probability~$\Phi_f(f_B)$ that a star with magnitude~$V_p$ will be blended with one or more physically unassociated stars that have total flux~$f_B$. If the primary star has magnitude outside the formal magnitude limits of the survey, it must be blended with at least one other star with magnitude inside the formal limits~(dashed lines). For primary stars with magnitude inside the formal limits, there is no such restriction~(solid lines).}
    \label{fig:fblends}
  \end{center}
\end{figure}

\begin{figure}
  \begin{center}
    \subfigure{
      \includegraphics[width=0.45\linewidth]{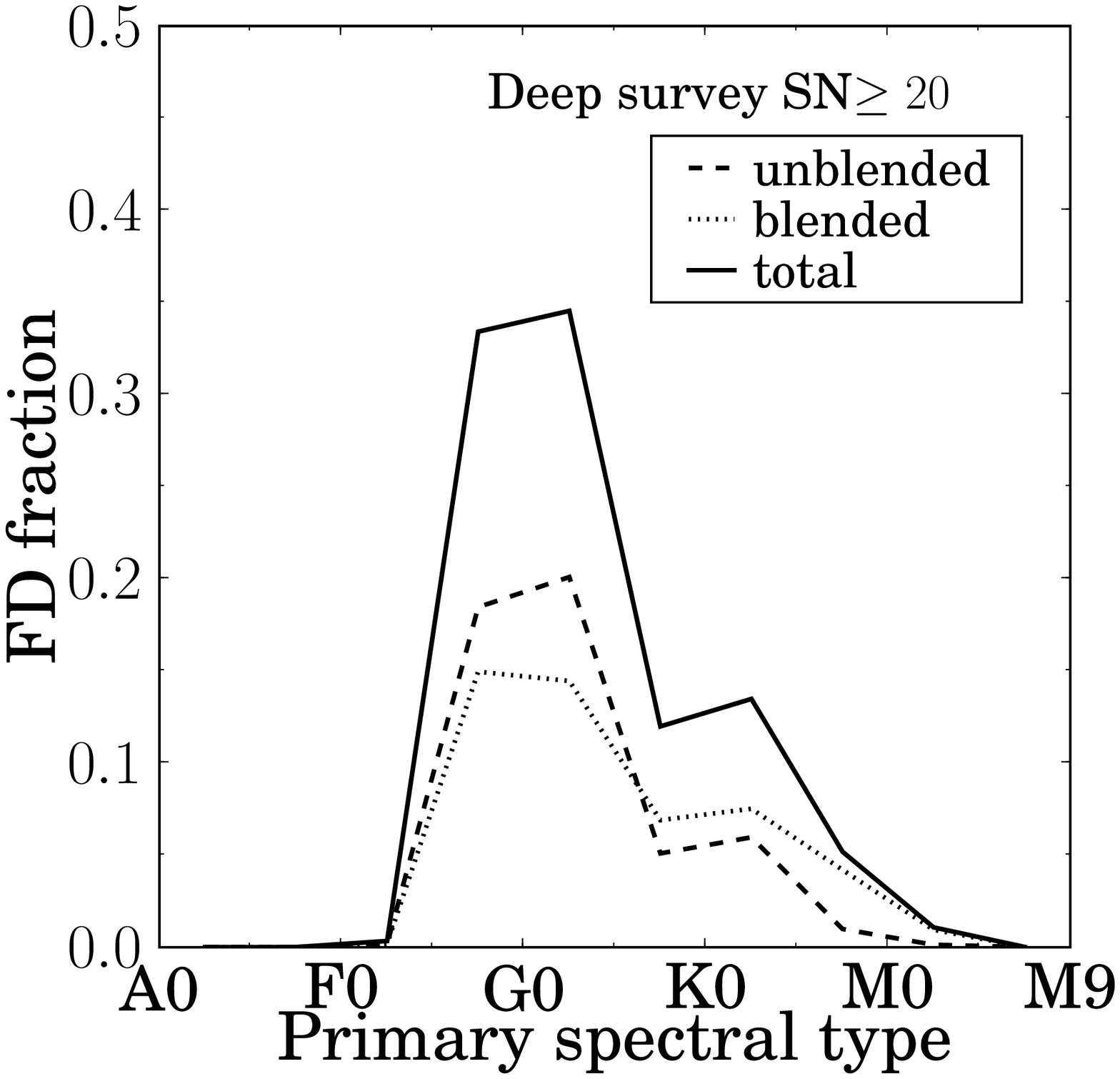}
      \label{}
      }
    \subfigure{
      \includegraphics[width=0.45\linewidth]{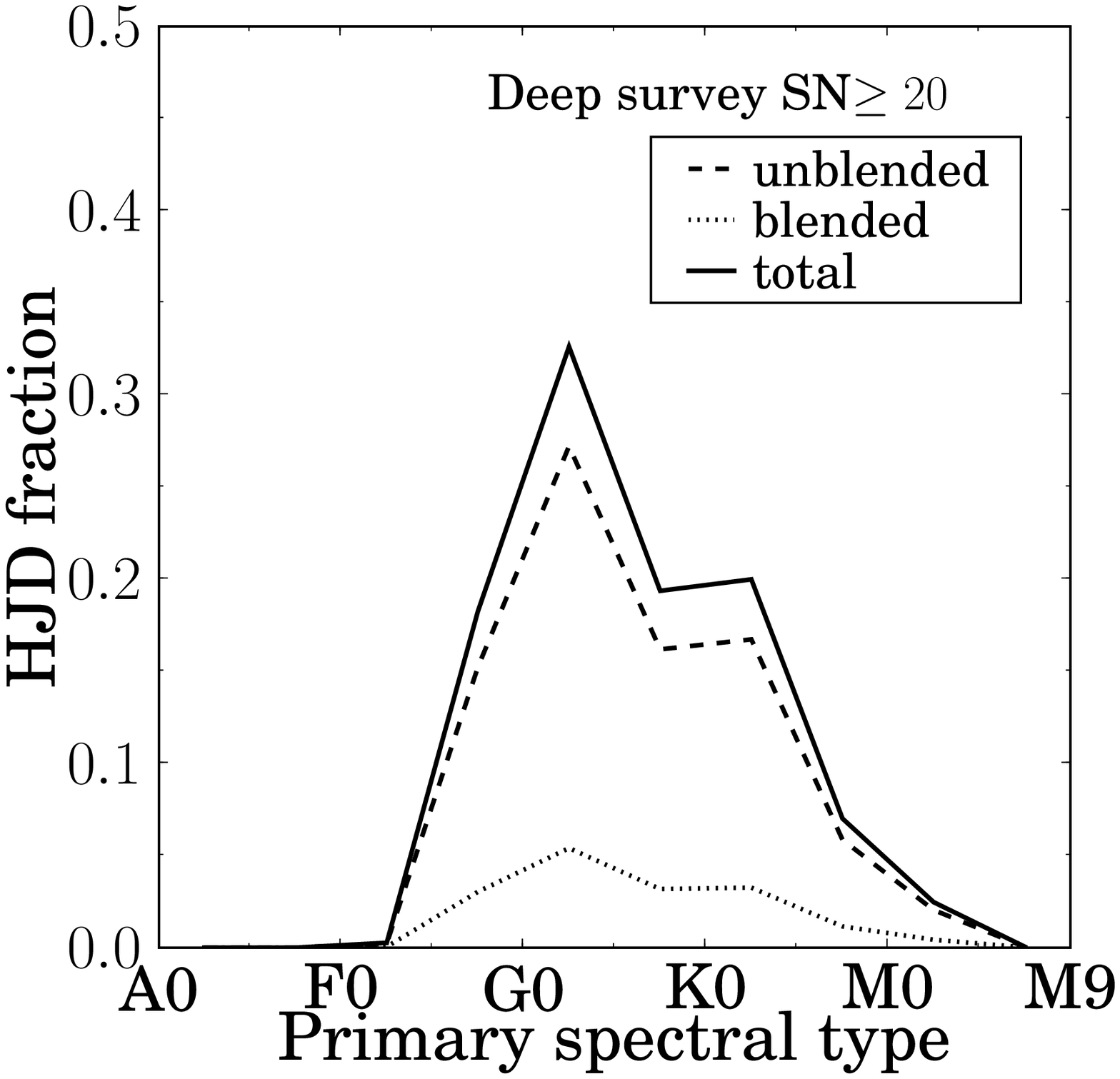}
      \label{}
      }
    \subfigure{
      \includegraphics[width=0.45\linewidth]{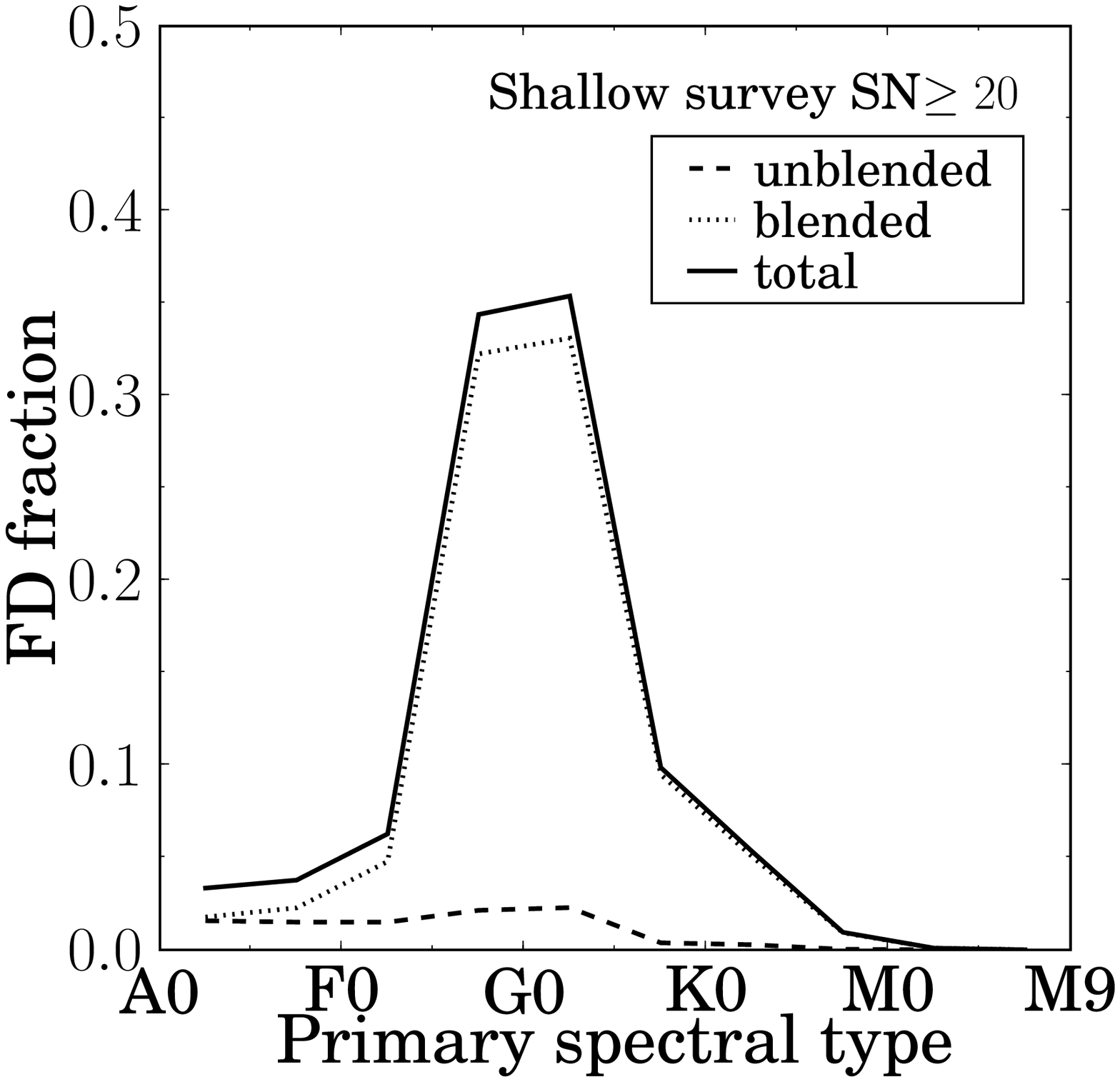}
      \label{}
      }
    \subfigure{
      \includegraphics[width=0.45\linewidth]{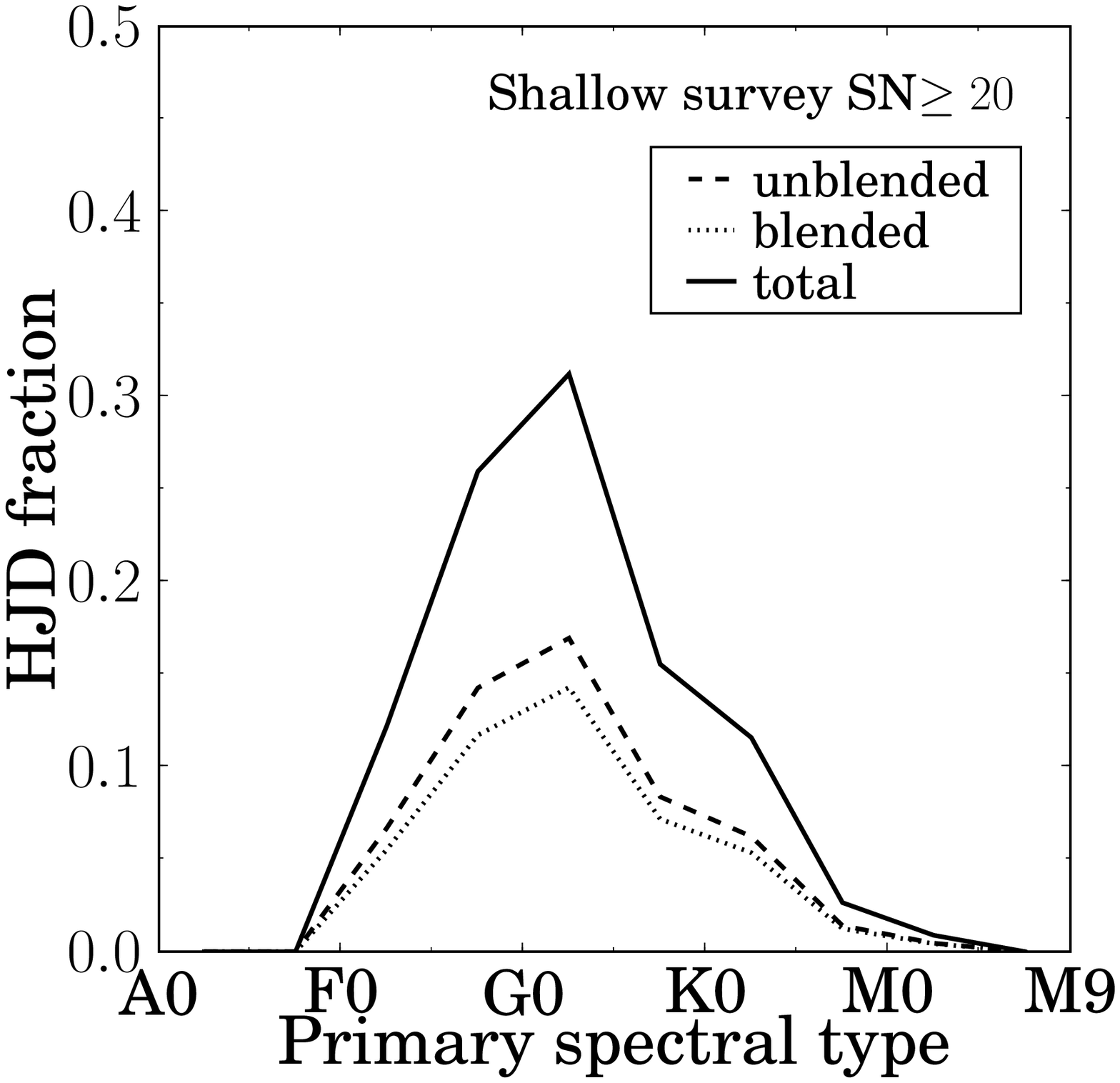}
      \label{}
      }
    \caption{Fraction of false detections~(FDs) in the deep (\emph{top left}) and shallow (\emph{bottom left}) surveys, and Hot Jupiter detections~(HJDs) in the deep (\emph{top right}) and shallow (\emph{bottom right}) surveys, binned according to the spectral type of the primary. Ten spectral bins are used. Results for blended (\emph{dotted}) and unblended (\emph{dashed}) configurations are plotted separately, as well as combined (\emph{solid line}). We only plot results for the limiting case that all tertiary components are resolved, as no significant qualitative or quantitative change is observed in the converse limiting case (see discussion in \S{\ref{sec:rates}}).  Results are shown for S/N$\geq20$.}
    \label{fig:res12}
  \end{center}
\end{figure}

\begin{figure}
  \begin{center}
    \subfigure{
      \includegraphics[width=0.45\linewidth]{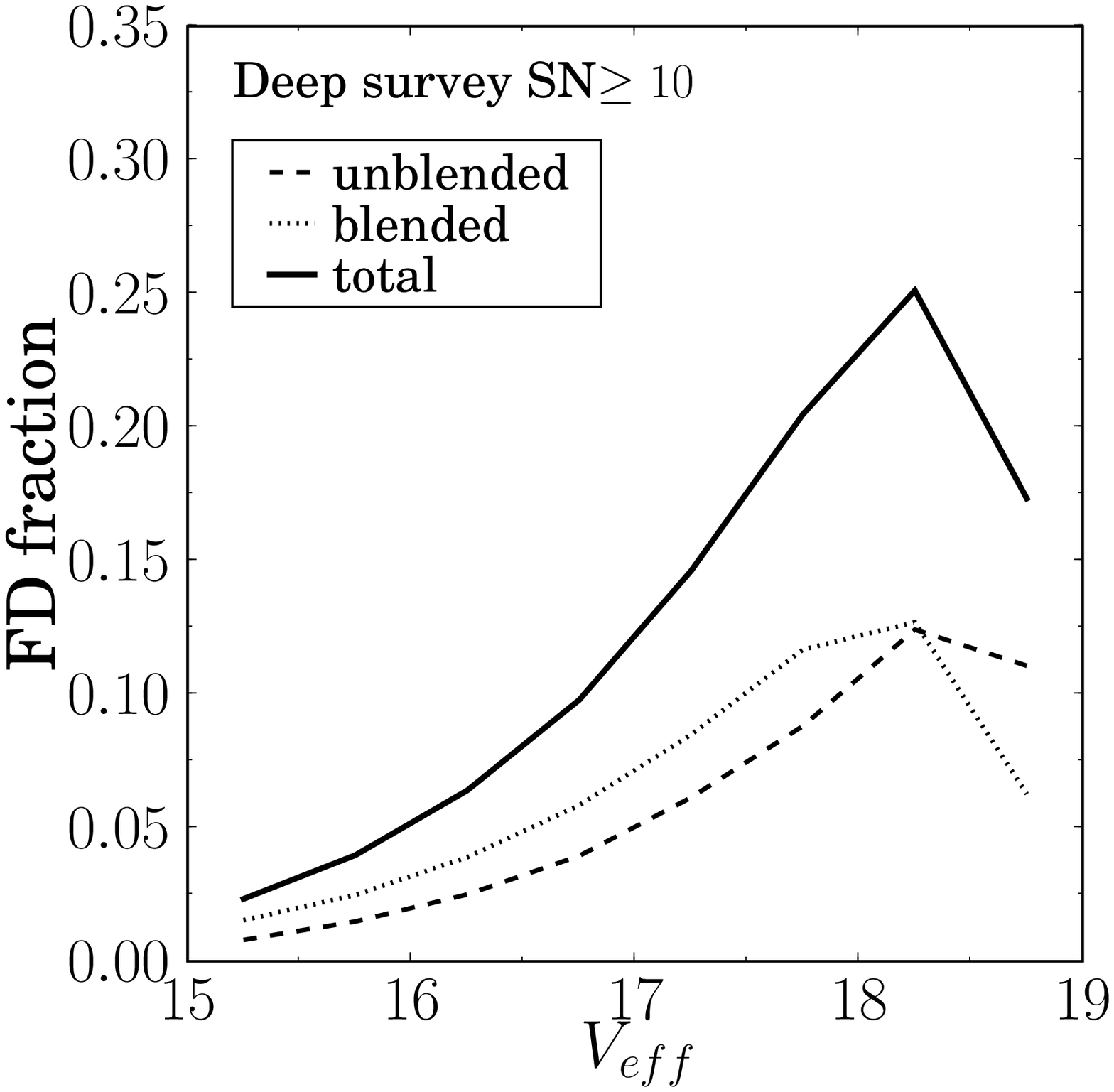}
      \label{}
      }
    \subfigure{
      \includegraphics[width=0.45\linewidth]{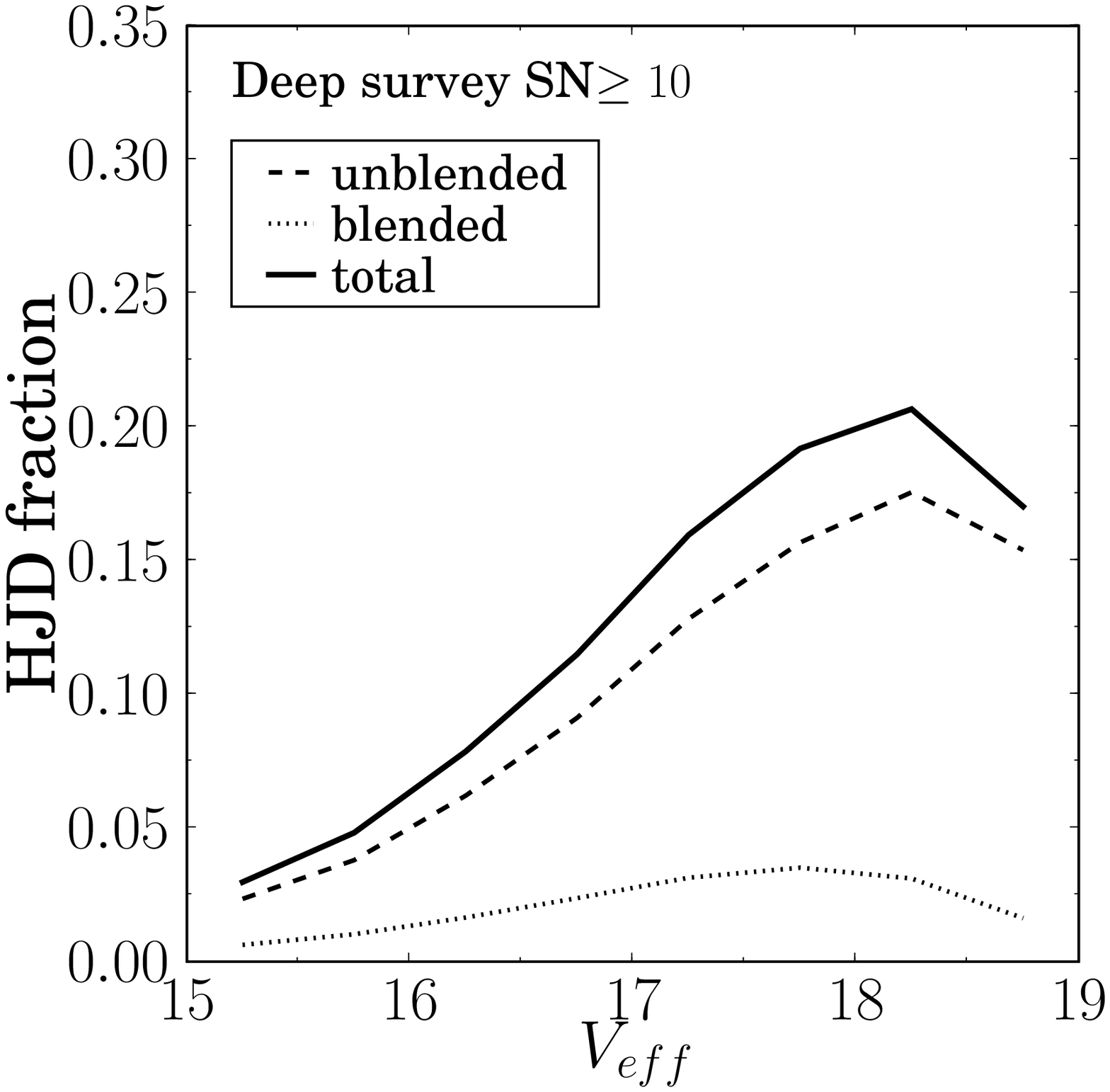}
      \label{}
      }
    \subfigure{
      \includegraphics[width=0.45\linewidth]{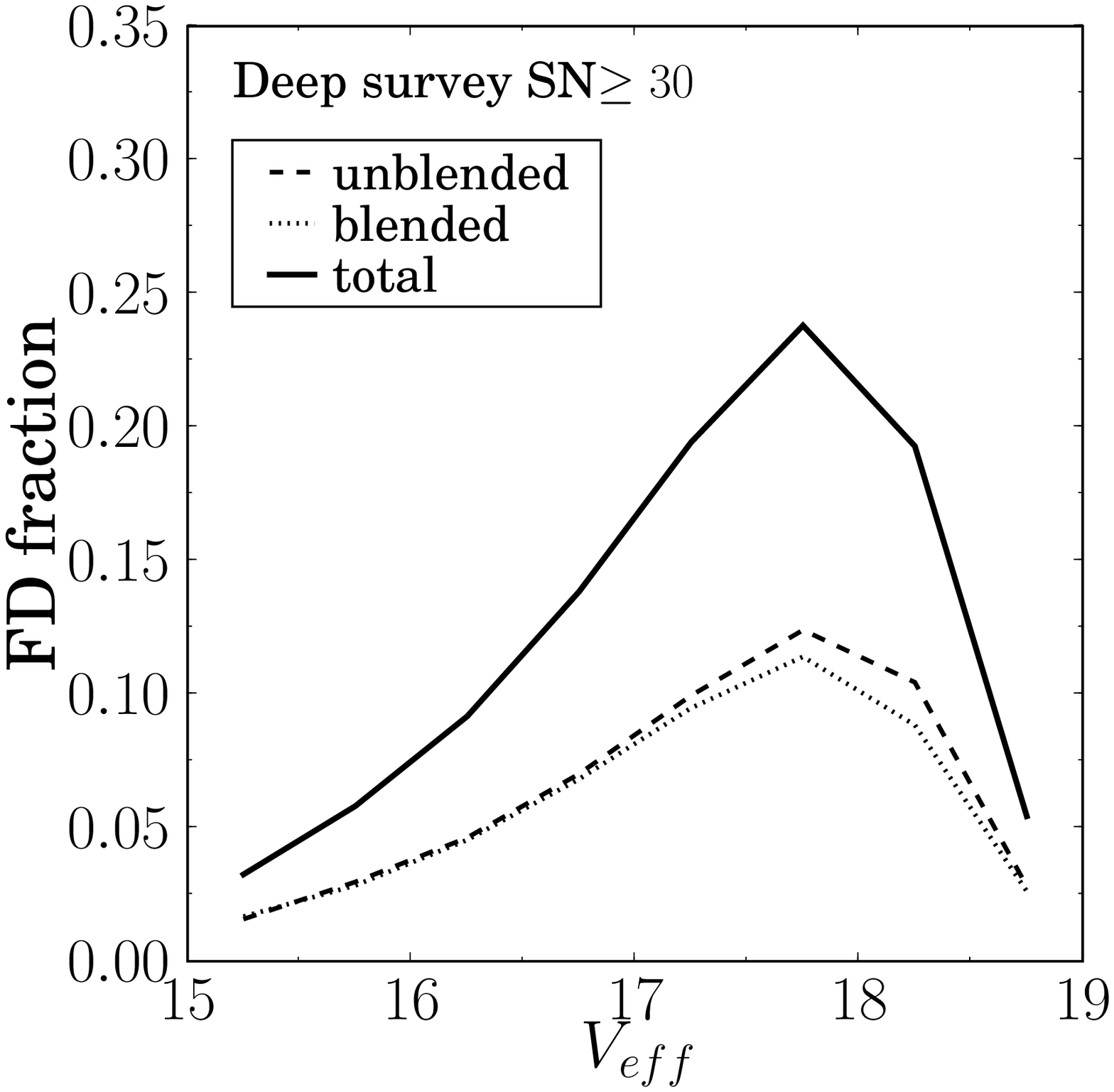}
      \label{}
      }
    \subfigure{
      \includegraphics[width=0.45\linewidth]{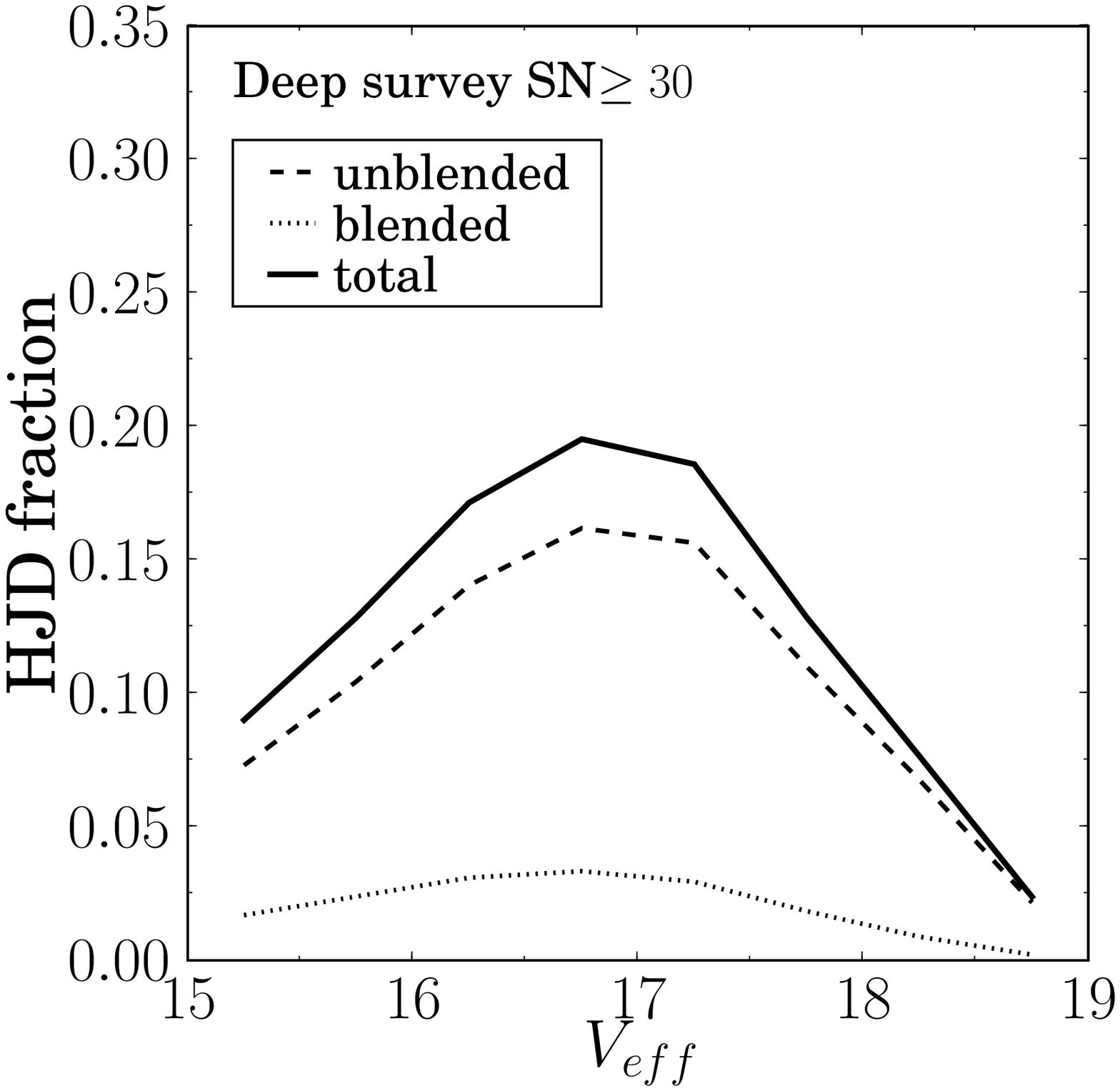}
      \label{}
      }
    \caption{As in Fig.~\ref{fig:res12}, but for the deep survey only and with detection fractions binned according to the effective magnitude, using a bin width of $0.5$~mag. Results are shown for S/N$\geq10$ and S/N$\geq30$ to illustrate how increasing the SN cutoff biases the detections towards brighter magnitudes (see text for discussion).}
    \label{fig:dres34}
  \end{center}
\end{figure}

\begin{figure}
  \begin{center}
    \subfigure{
      \includegraphics[width=0.45\linewidth]{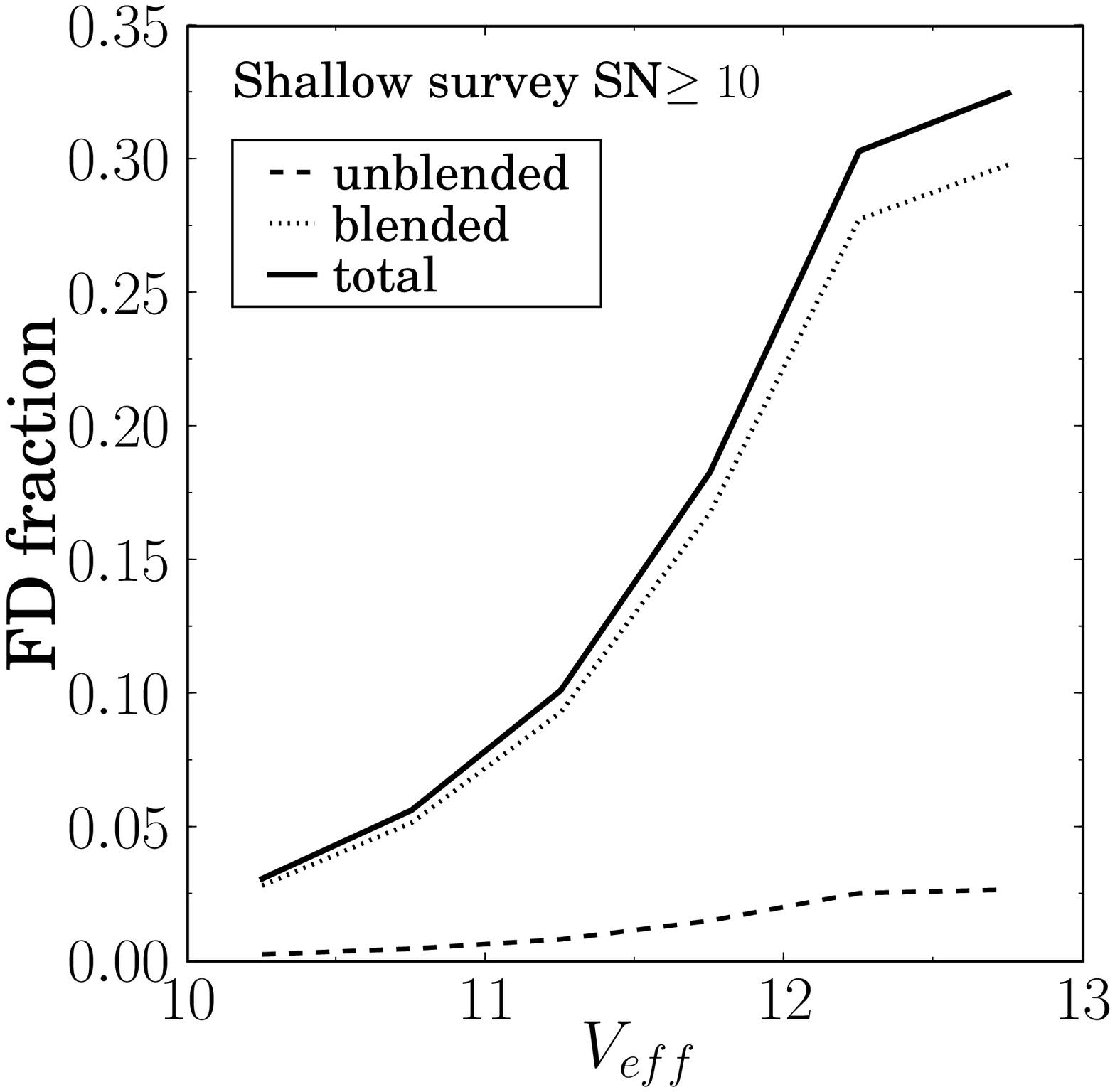}
      \label{}
      }
    \subfigure{
      \includegraphics[width=0.45\linewidth]{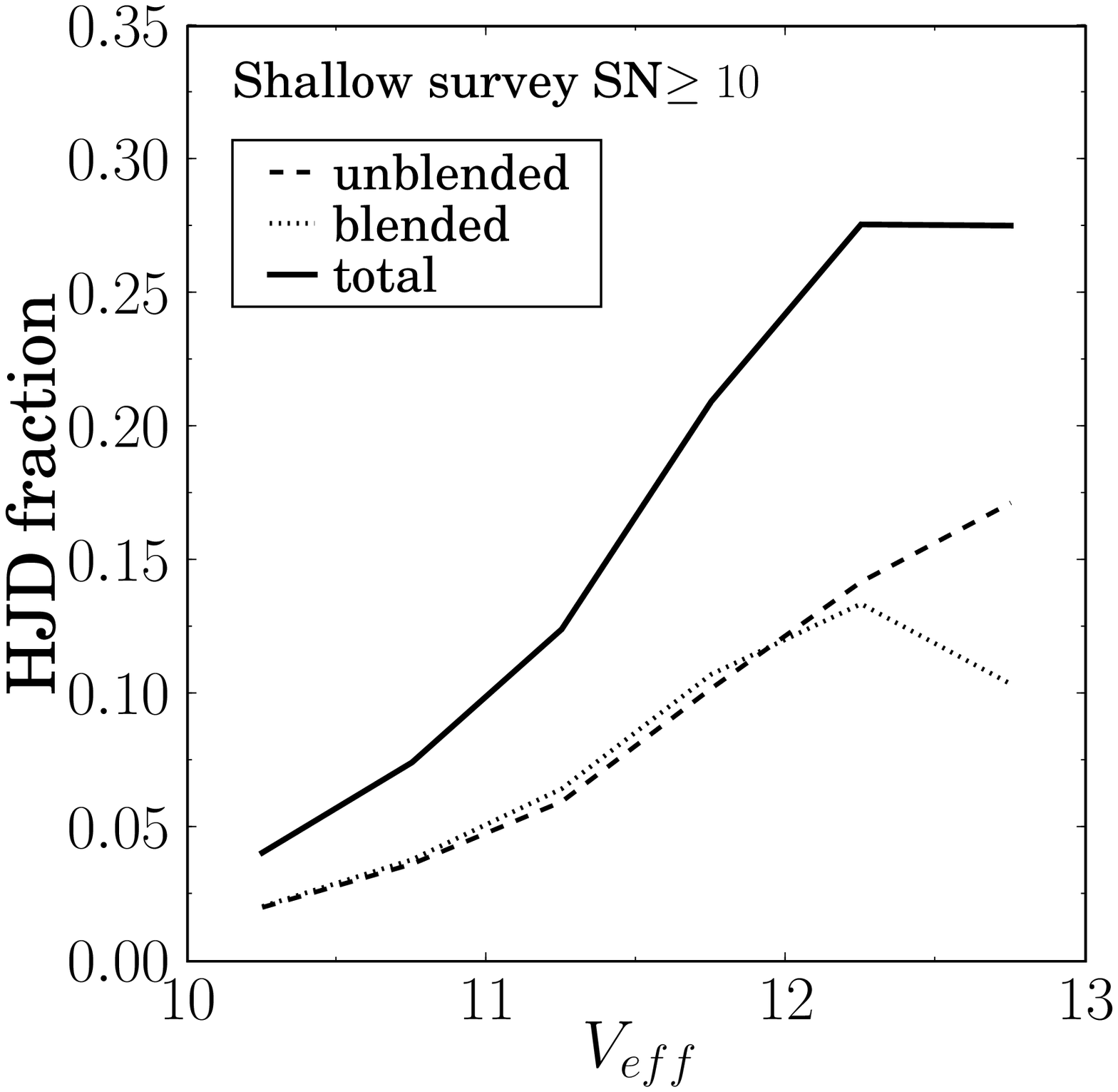}
      \label{}
      }
    \subfigure{
      \includegraphics[width=0.45\linewidth]{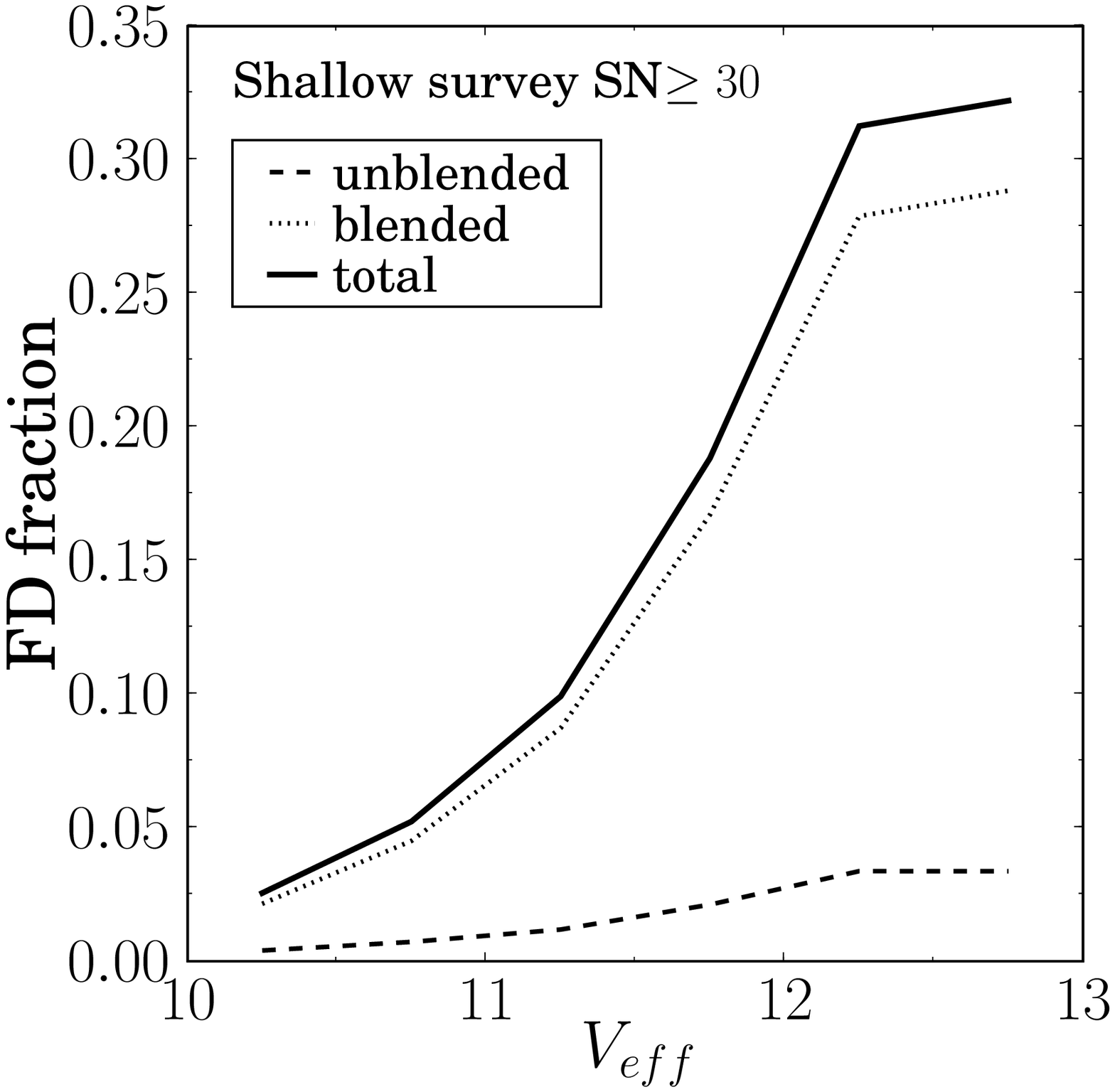}
      \label{}
      }
    \subfigure{
      \includegraphics[width=0.45\linewidth]{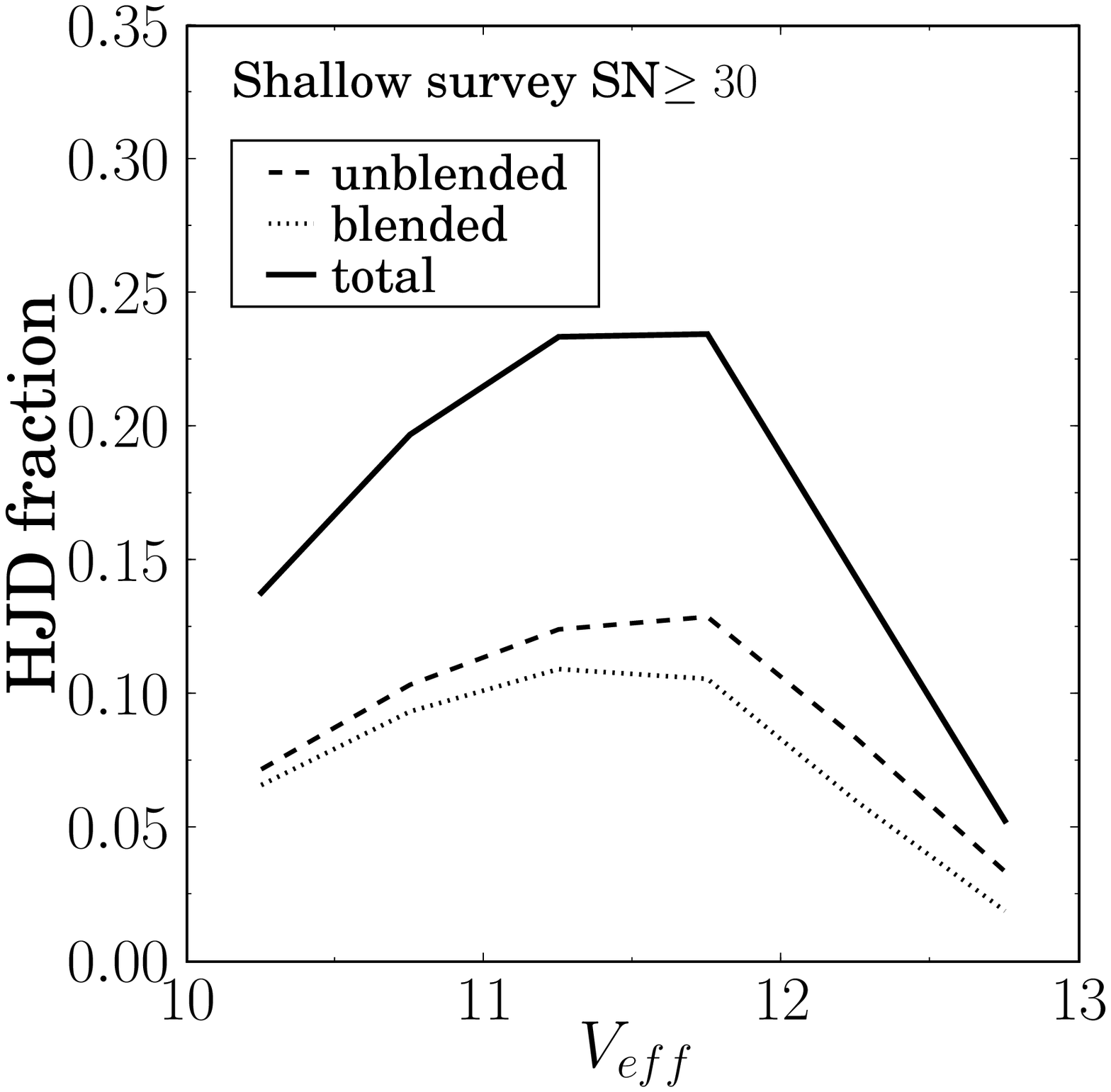}
      \label{}
      }
    \caption{As in Fig.~\ref{fig:dres34}, but for the shallow survey.}
    \label{fig:sres34}
  \end{center}
\end{figure}

\begin{figure}
  \begin{center}
    \subfigure{
      \includegraphics[width=0.45\linewidth]{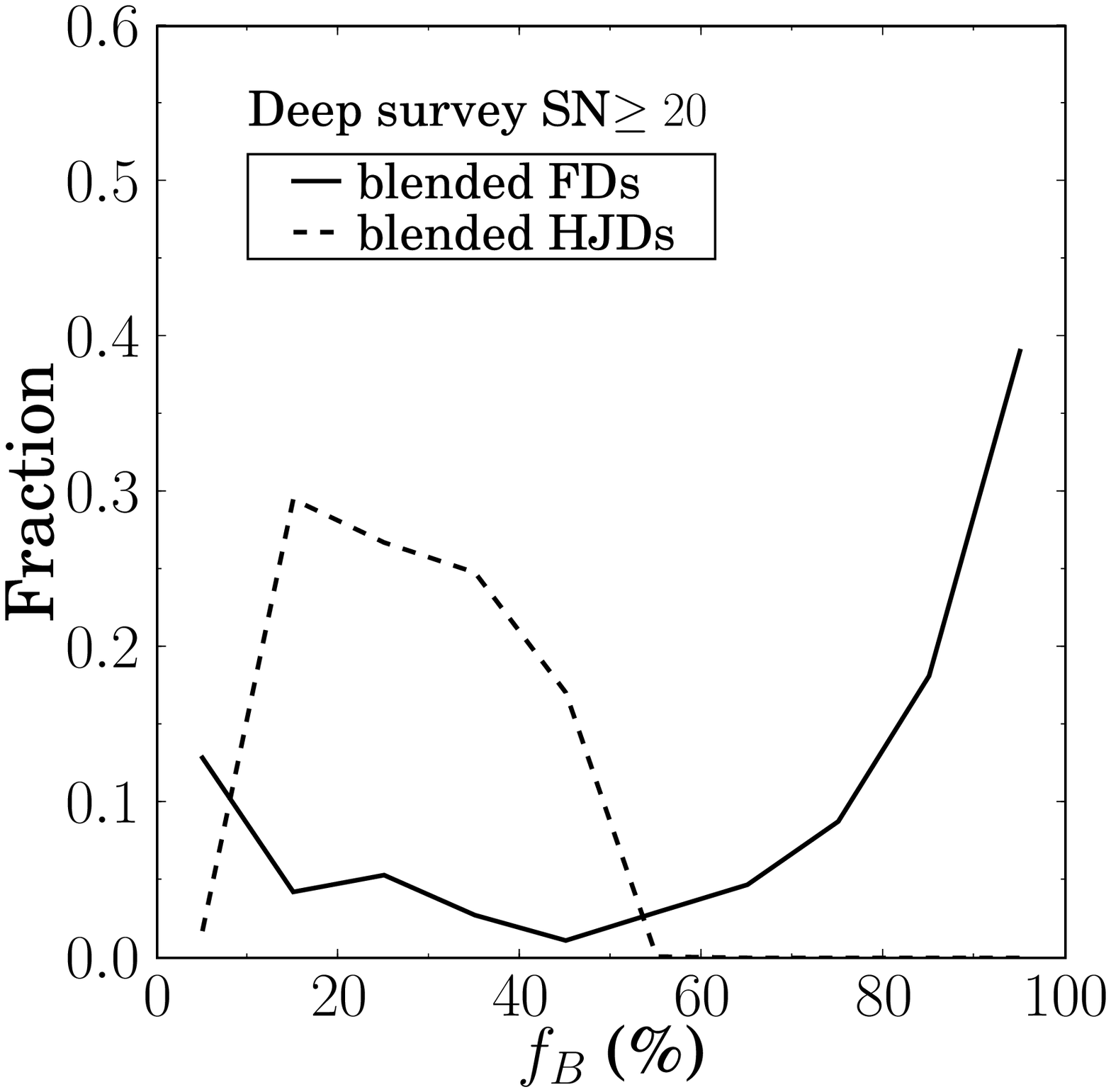}
      \label{}
      }
    \subfigure{
      \includegraphics[width=0.45\linewidth]{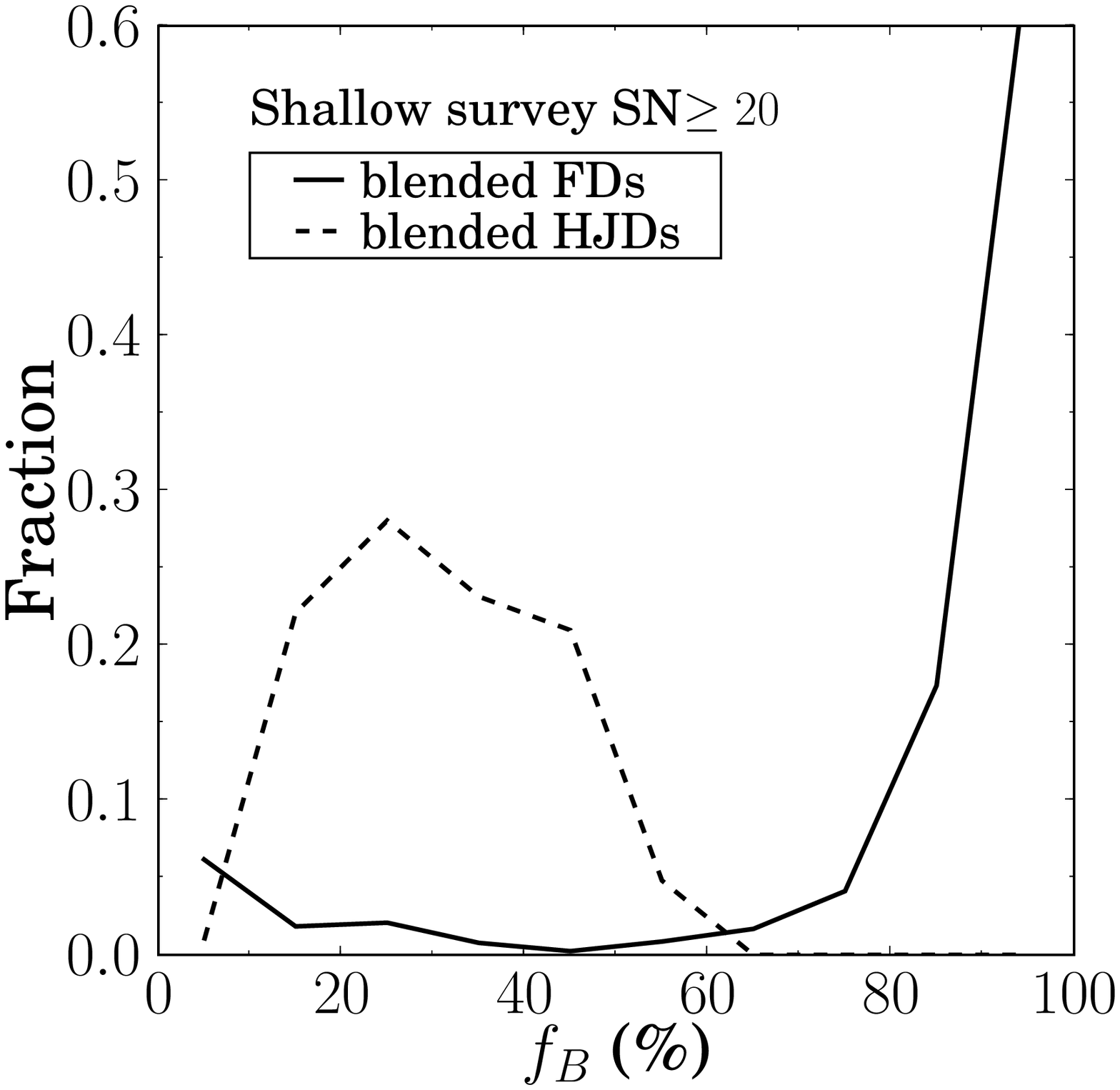}
      \label{}
      }
    \caption{Detection fractions for blended configurations, binned according to the percentage of the total flux that is due to blended stars, using a bin width of $10\%$.}
    \label{fig:res56}
  \end{center}
\end{figure}

\begin{figure}
  \begin{center}
    \subfigure{
      \includegraphics[width=0.45\linewidth]{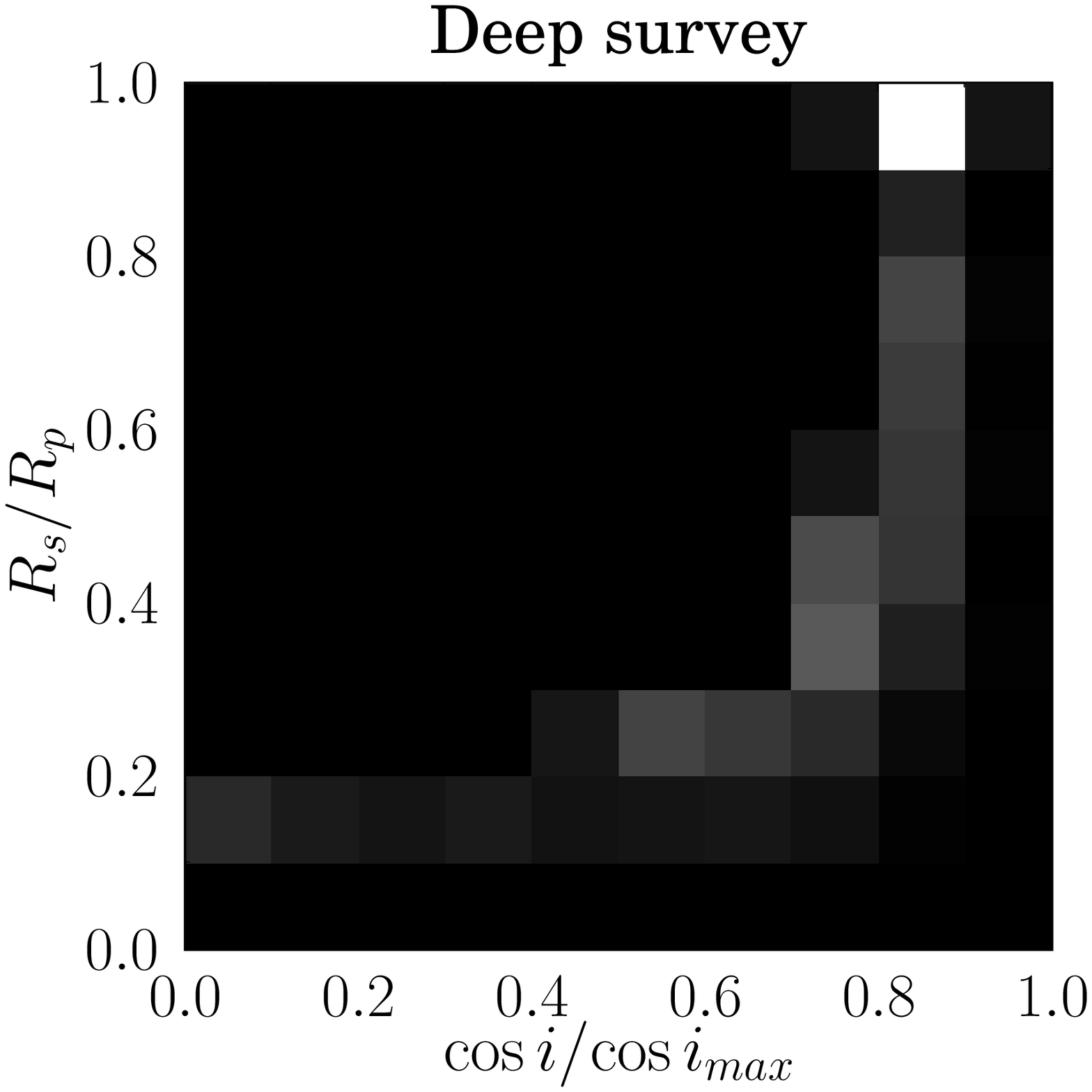}
      \label{}
      }
    \subfigure{
      \includegraphics[width=0.45\linewidth]{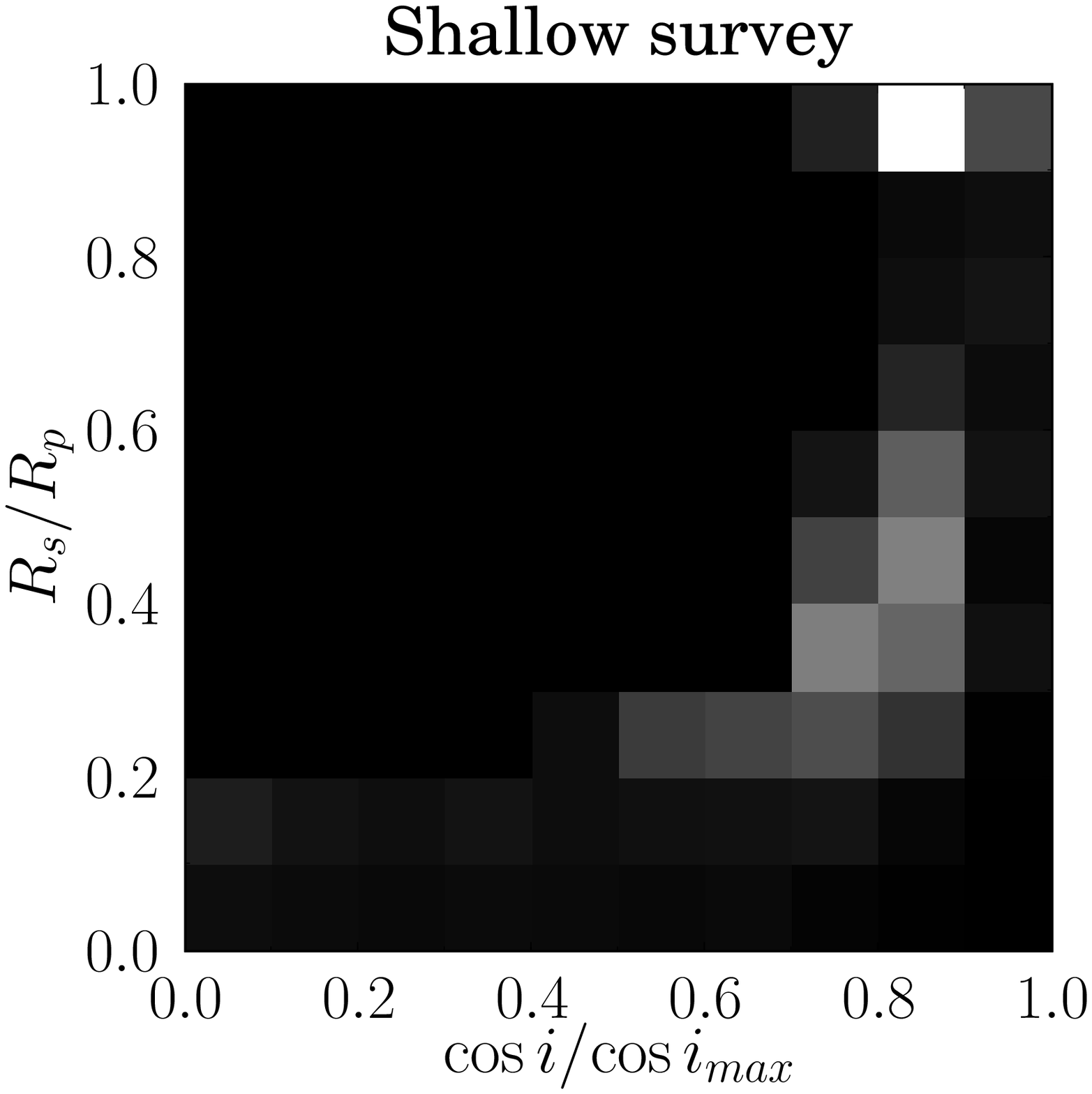}
      \label{}
      }
    \caption{Variation of unblended false detection rates in the deep survey (\emph{left}) and in the shallow survey (\emph{right}), with respect to transit latitude and the relative size of the eclipsing components, $R_s/R_p$. On the grayscale colormap, white corresponds to relatively high detection rates and black corresponds to relatively low detection rates. Results are shown for S/N$\geq20$ and are effectively identical for both surveys, with most detections caused by high-latitude transits. In particular, twins~($R_p  \approx R_s$) undergoing high-latitude eclipses are the most populous category of false detections. These configurations will produce indistinguishable primary and secondary transits that are shallow enough to be mistaken for a Hot Jupiter transit.}
    \label{fig:res7}
  \end{center}
\end{figure}

\begin{figure}
  \begin{center}
    \subfigure{
      \includegraphics[width=0.45\linewidth]{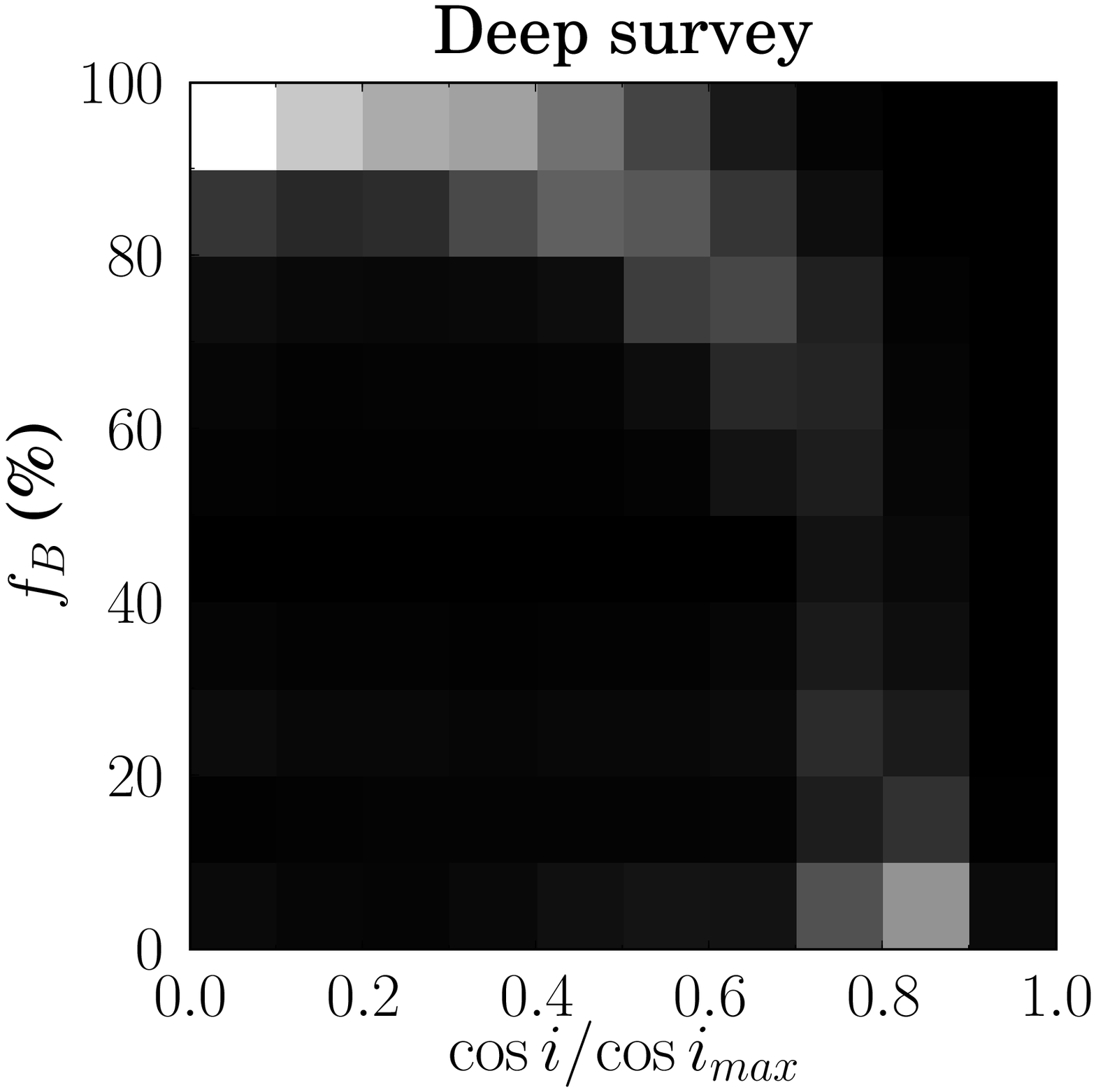}
      \label{}
      }
    \subfigure{
      \includegraphics[width=0.45\linewidth]{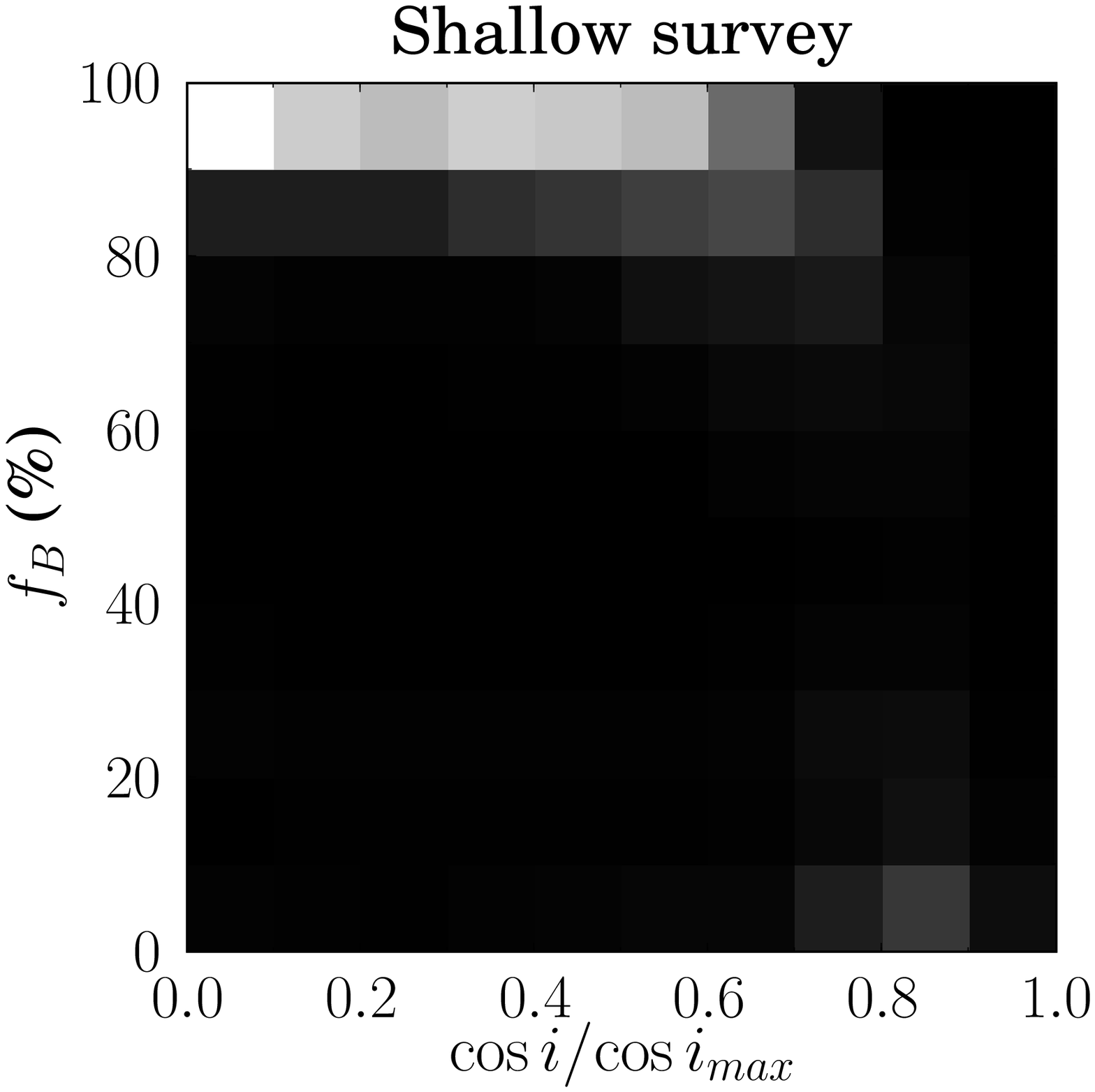}
      \label{}
      }
    \caption{Variation of blended false detection rates in the deep survey (\emph{left}) and in the shallow survey (\emph{right}), with respect to transit latitude and the blended flux expressed as a percentage of the total flux. Grayscale is the same as in Fig.~\ref{fig:res7} and results are again shown for S/N$\geq20$. The most populous category of false detections is those for which the blended flux contributes more than $80\%$ of the total flux. A smaller peak is also observed for detections in which the blended flux contributes less than $20\%$ of the total flux and the transit is high-latitude.}
    \label{fig:res8}
  \end{center}
\end{figure}

\begin{deluxetable}{lcc}  
\tabletypesize{\scriptsize}
\tablecaption{Properties of the deep and shallow synthetic surveys\label{tab:survs}}
\tablewidth{0pt}
\tablehead{
\colhead{} & \colhead{Deep} & \colhead{Shallow}
}
\startdata
 FOV longitude~($l$) & $331.5^{\circ}$ & $331.5^{\circ}$ \\
 FOV latitude~($b$) & $11.0^{\circ}$ & $11.0^{\circ}$ \\
 Formal~mag.~range & $15<V<19$ & $10<V<13$ \\
 Extended~mag.~range & $15<V<22$ & $10<V<16$ \\
 Confusion radius~($r_c$) & $2''$ & $20''$ \\
 Maximum period~($P$) & 10\,d & 10\,d \\
 Parameters for photometric & & \\
 scatter function\tablenotemark{\dagger}: & & \\
 \multicolumn{1}{c}{$a$} & $+0.018$ & $+0.029$  \\
 \multicolumn{1}{c}{$b$} & $-0.357$ & $-0.477$  \\
 \multicolumn{1}{c}{$c$} & $-1.161$ & $-0.726$  \\
\enddata
\tablenotetext{\dagger}{$\log_{10} \sigma = a V^2 + b V + c$ \hspace{0.5cm} (see \S\ref{sec:configs})}
\end{deluxetable}

\begin{deluxetable}{ccc}  
\tabletypesize{\scriptsize}
\tablecaption{Probability $p_n$ of $n$ blended stars \label{tab:nblends}}
\tablewidth{0pt}
\tablehead{
\colhead{} & \colhead{Deep survey} & \colhead{Shallow survey}
}
\startdata
 $n$ & $p_n$ & $p_n$ \\
 0 & 0.818 & 0.528 \\
 1 & 0.164 & 0.337 \\
 2 & 0.017 & 0.108 \\
 3 & 0.001 & 0.023 \\
 4 & 0.000 & 0.004 \\
\enddata
\end{deluxetable}


\begin{deluxetable}{clcclccc}  
\tabletypesize{\scriptsize}
\tablecaption{Estimated rates of false detections and Hot Jupiter detections per 10,000 resolved stars monitored assuming that all tertiary components are resolved\tablenotemark{\star}    \label{tab:tcrrates}}
\tablewidth{0pt}
\tablehead{
\colhead{Deep survey} & &&&
}
\startdata
             & \multicolumn{2}{l}{False detections:} && \multicolumn{2}{l}{Hot Jupiter detections:} &&   \\
 S/N$\,\geq$ & \multicolumn{1}{l}{Blended}   & Unblended & Total       &  \multicolumn{1}{l}{Blended}    & Unblended   & Total       & False/Planet ratio   \\
 10          & 2.0 (2.4)  & 1.8 (1.9) & 3.8 (4.3) &  0.55 (0.69) & 2.7 (3.4)   & 3.3 (4.1) &  1.2 (1.1)  \\           
 20          & 1.4 (1.6)  & 1.4 (1.6) & 2.8 (3.2) &  0.26 (0.32) & 1.3 (1.6)   & 1.6 (1.9) &  1.8 (1.7)  \\
 30          & 0.96 (1.1) & 1.0 (1.1) & 2.0 (2.2) &  0.14 (0.17) & 0.71 (0.85) & 0.85 (1.0) &  2.3 (2.2)  \\
 \cutinhead{Shallow survey}
             & \multicolumn{2}{l}{False detections:} && \multicolumn{2}{l}{Hot Jupiter detections:} &&   \\
 S/N$\,\geq$ & \multicolumn{1}{l}{Blended}   & Unblended & Total  & \multicolumn{1}{l}{Blended}      & Unblended & Total     & False/Planet ratio: \\
 10          & 3.6 (4.4)  & 0.33 (0.37) & 3.9 (4.8) & 0.30 (0.37)   & 0.34 (0.42)   & 0.64 (0.79) & 6.1 (6.0) \\
 20          & 3.0 (3.6)  & 0.34 (0.38) & 3.3 (4.0) & 0.12 (0.14)   & 0.14 (0.17)   & 0.26 (0.31) & 13 (13) \\
 30          & 2.5 (3.0)  & 0.32 (0.36) & 2.8 (3.4) & 0.054 (0.063) & 0.064 (0.076) & 0.12 (0.14) & 24 (24) \\
\enddata
\tablenotetext{\star}{Values without parentheses give the predicted detection rates with windowing effects accounted for and the requirement that at least three transits are observed, while values in parentheses give the detection rates ignoring any windowing effects.}
\end{deluxetable}

\begin{deluxetable}{clclccc}  
\tabletypesize{\scriptsize}
\tablecaption{Estimated rates of false detections  per 10,000 resolved stars monitored assuming that all tertiary components are unresolved\tablenotemark{\ddagger}    \label{tab:tcurates}}
\tablewidth{0pt}
\tablehead{
\colhead{Deep survey} & &&&&&
}
\startdata
             & \multicolumn{2}{l}{Doubles:} & \multicolumn{2}{l}{Triples:} &&  \\
 S/N$\,\geq$ & \multicolumn{1}{l}{Blended}     & Unblended          & \multicolumn{1}{l}{Blended}    & Unblended  & Total & False/Planet ratio  \\
 10          & 0.34 (0.51) & 0.24 (0.32)  & 1.6 (1.9)   & 1.5 (1.6)  & 3.7 (4.2) &  1.1 (1.1) \\           
 20          & 0.23 (0.34) & 0.21 (0.27)  & 1.2 (1.3)   & 1.3 (1.3)  & 2.8 (3.3) &  1.9 (1.7)  \\
 30          & 0.15 (0.21) & 0.14 (0.19)  & 0.85 (0.93) & 0.96 (1.0) & 2.1 (2.4) &  2.5 (2.3)  \\
 \cutinhead{Shallow survey}
            & \multicolumn{2}{l}{Doubles:} & \multicolumn{2}{l}{Triples:} &&   \\
 S/N$\,\geq$ & \multicolumn{1}{l}{Blended}    & Unblended           &  \multicolumn{1}{l}{Blended}  & Unblended   & Total & False/Planet ratio   \\
  10         & 0.66 (0.94) & 0.048 (0.061) &  2.9 (3.3) & 0.25 (0.27) & 3.9 (4.6) &  6.0 (5.8)  \\           
 20          & 0.53 (0.75) & 0.051 (0.064) &  2.5 (2.8) & 0.26 (0.28) & 3.3 (3.9) &  13 (13)  \\
 30          & 0.44 (0.61) & 0.051 (0.065) &  2.1 (2.3) & 0.27 (0.29) & 2.9 (3.3) &  24 (24)  \\
\enddata
\tablenotetext{\ddagger}{Rates presented as in Table~\ref{tab:tcrrates}. Hot Jupiter detection rates are not shown as they do not change from Table~\ref{tab:tcrrates}.}
\end{deluxetable}

\end{document}